\documentclass[aoas,preprint]{imsart}
\RequirePackage{amsthm,amsmath,amsfonts,amssymb}
\usepackage{graphicx}
\usepackage{subcaption}
\usepackage{bm}
\usepackage[hidelinks]{hyperref}
\usepackage[authoryear]{natbib}

\begin{document}

\begin{frontmatter}
\title{Hierarchical Multidimensional Scaling for the Comparison of Musical Performance Styles}
\runtitle{HMDS}

\begin{aug}

\author{\fnms{Anna K.} \snm{Yanchenko}\ead[label=e2,mark]{anna.yanchenko@duke.edu}}
\and
\author{\fnms{Peter D.} \snm{Hoff}\ead[label=e3,mark]{peter.hoff@duke.edu}}

\address{Department of Statistical Science,
Duke University,
\printead{e2,e3}}
\end{aug}

\begin{abstract}
Quantification of stylistic differences between musical artists is of academic interest to the music community, and is also useful for other applications such as music information retrieval and recommendation systems. Information about stylistic differences can be obtained by comparing the performances of different artists across common musical pieces. In this article, we develop a statistical methodology for identifying and quantifying systematic stylistic differences among artists that are consistent across audio recordings of a common set of pieces, in terms of several musical features. Our focus is on a comparison of ten different orchestras, based on data from audio recordings of the nine Beethoven symphonies. As generative or fully parametric models of raw audio data can be highly complex, and more complex than necessary for our goal of identifying differences between orchestras,  we propose to reduce the data from a set of audio recordings down to pairwise distances between orchestras, based on different musical characteristics of the recordings, such as tempo, dynamics, and timbre.  For each of these characteristics, we obtain multiple pairwise distance matrices, one for each movement of each symphony. We develop a hierarchical multidimensional scaling (HMDS) model to identify and quantify systematic differences between orchestras in terms of these three musical characteristics, and interpret the results in the context of known qualitative information about the orchestras. This methodology is able to recover several expected systematic similarities between orchestras, as  well as to identify some more novel results.  For example, we find that modern recordings exhibit a high degree of similarity to each other, as compared to older recordings. 
\end{abstract}


\end{frontmatter}

\section{Introduction}\label{sec:intro}

The quantification of stylistic differences between musical artists 
is of interest in musicology, 
and has uses for the
general music-listening public, such as for  music information retrieval and recommender systems. In this article, we are particularly focused on 
the quantification of systematic
differences among ten different orchestras, 
based on data from audio recordings of the nine Beethoven symphonies.
This is motivated by a desire to statistically quantify a variety of 
descriptions of heterogeneity among orchestras that has typically been done 
qualitatively, such as how
musical performances might change over time, or how 
orchestras from the United States 
systematically differ from European orchestras. 

In general, information about stylistic differences among artists can be obtained by comparing their performances of a common collection of musical pieces. Quantitative analyses of different orchestral recordings has been explored in the music information retrieval community using tempo curve analysis \citep{PeperkampEtAl:2017} and image analysis techniques with principal components analysis \citep{LiemHanjalic:2015}. A more statistical approach to making such comparisons would be to fit a model to the audio data for each artist separately, and then compare the estimated model parameters corresponding to each artist. However, from a data analysis perspective, an audio recording of a piece of music is a complex, multivariate, highly structured time series, with long-term dependencies that, in symphonic works, exist over multiple minutes.  

Comparison is a critical question of interest for musicologists \citep{Cook:2005}, and a quantitative comparison of systematic similarities across performances is of interest to the musicology and music information retrieval communities in several regards.  First, such an approach allows for the quantitative comparison of a large corpus of musical pieces, for a recommendation system \citep{NIPS2013_5004} or music corpus analysis task \citep{LiemHanjalic:2015} in the music information retrieval context.  Second, such an approach is of interest for musicologists as well \citep{Cook:2005}, since it allows for numerically exploring, and testing, hypotheses of interest, such as whether more recent recordings exhibit less variation than older recordings \citep{Liebman:2012}, a primary question of interest in this work, as well.  Finally, this type of analysis can serve as a starting point for insight into understanding differences in expressive playing from a psychological and cognition perspective \citep{Desain:1994, Penel:1998, Goebl:2009, Liem:2011, Liem:2011aa}.  While the complexity of audio data allows for analysis of more nuanced features such as musical style, this complexity also makes it challenging to develop accurate generative statistical models of musical audio data in its raw form. 
One popular approach is to use convolutional neural networks with dilated, causal convolutions \citep{van-den-Oord:EtAl:2016}, and while these can generate audio that mimics their input, the large number of estimated parameters can be difficult to interpret, and potentially non-comparable across model fits.  Additionally, WaveNet \citep{van-den-Oord:EtAl:2016} models very short audio segments of only a few seconds in length, making stylistic analysis over full orchestral works that are multiple minutes in length challenging. 

As an alternative to such generative approaches, for the purpose of identifying differences between orchestras, we propose to reduce the data from a set of audio recordings down to pairwise distances between orchestras, based on different musical characteristics of the recordings, such as tempo, volume dynamics, and timbre.  
For each of these characteristics, we obtain multiple pairwise distance matrices, one for each movement of each symphony, resulting in 37 distance matrices for each of the 3 musical characteristics. Comparison of the orchestras may then proceed using statistical methods appropriate for analysis of distance data.  Such methods might include distance-based analysis of variance (ANOVA) approaches used in the ecological community \citep{Anderson:2001, McArdle-Anderson:2001, Minas:2014, Rizzo:2010}, or the related functional ANOVA (FANOVA) approach that was developed to analyze distance data in genomics with Gaussian processes \citep{Vsevolozhskaya:2014}. In this article, we focus on adapting multidimensional scaling (MDS)  to the specific task of combining information across multiple distance matrices to identify consistent differences between orchestras. MDS is a popular technique for analyzing distance data, originally developed in the psychology literature \citep{Torgerson:1952}.  Standard MDS generates an embedding of observed distance data into a Euclidean space so that the distances between objects in the embedding approximates the observed distances.

Our HMDS model can be viewed as a modification and extension of the Bayesian MDS model proposed by \citet{Oh-Raftery:2001}. In Bayesian MDS (BMDS), the observed distance matrix is assumed to be equal to the distance matrix of a set of latent vectors in a Euclidean space, plus (truncated) Gaussian noise. Using a Markov chain Monte Carlo approximation algorithm, the posterior distribution of the latent vectors may be inferred from an observed matrix of distance data. 
Our proposed HMDS extends the BMDS model of \citet{Oh-Raftery:2001} in several ways to accommodate specific features of our data. 
Most importantly, BMDS was developed to analyze a single distance matrix, and assumes that the ``true'' distances are Euclidean. In contrast, our data consist of 
37 distance values for each pair of orchestras and each musical characteristic. 
Treating these 37 values as ``replicates'' our HMDS model is able to distinguish between differences that are 
``systematic'', i.e.\ consistent across musical pieces, and differences that are idiosyncratic to particular pieces. Furthermore, our approach allows for the systematic  component of the observed distance matrices to be non-Euclidean. This is useful, as the distance metrics we use to evaluate 
stylistic differences are not necessarily embeddable in Euclidean space. 
Another feature of our model is that we allow for differences in the potential for variation across replicates, or pieces. 
This is critical for our application, as 
some musical pieces have much more potential for variation than others. Combining information across  pieces without adjusting for this potential would tend 
to hide systematic effects. 
Finally, 
in contrast to the truncated normal model in BMDS,
we model non-negative distances using gamma distributions. 
This approach has the advantage of being able to 
accommodate skew in the distribution of observed distances, 
and is perhaps a more natural choice for positive distance data. 
Additionally, 
our gamma model for observed distances permits the use of semi-conjugate 
prior distributions, which facilitates several posterior calculations. 

Related to our approach, \citet{Park:2008} extended the BMDS model to multiple distance matrices in the specific context of capturing two types of heterogeneity in preference data.  Their model combined two major types of latent utility models for preference data in a generalized, mixture model framework.  However, in contrast to our work, \citet{Park:2008} specified normally distributed likelihood functions for their observed dominance score matrices that were specific to the two latent utility functions considered, as opposed to the general setting of skewed, positive distance data, as we consider here.  Additionally, the model in \citet{Park:2008} specified the same variance across the multiple distance matrices and assumed Euclidean distances between embedding vectors, while HMDS allows for heterogeneous potential for variation for each distance matrix and relaxes the Euclidean assumption.  Additional extensions to BMDS include Bayesian MDS with variable selection  \citep{Lin:2019}, which incorporated covariate information in the dimensionality reduction performed by classical MDS and allowed for heterogeneous variability by distance pair, and Bayesian MDS with simultaneous variable selection with dimension reparameterization \citep{Fong:2015}.  In contrast to HMDS, though, both \citet{Lin:2019} and \citet{Fong:2015} did not extend BMDS to multiple distance matrices and again assumed normal distributions for the observed distance data.  In summary, our proposed HMDS model is distinguished from prior work such as \citet{Park:2008, Lin:2019, Fong:2015} by the extension to multiple distance matrices with heterogeneous variation, the modeling of non-negative distances with gamma distributions and the relaxation of the Euclidean assumption of the systematic component of our observed distance matrices.

HMDS also extends previous approaches to the comparison of audio recordings \citep{PeperkampEtAl:2017, LiemHanjalic:2015, Sapp:2007, Sapp:2008, Liebman:2012} from the perspective of the musical application of interest. HMDS seeks to find systematic similarities across recordings in a comparative and unsupervised way, and as such, does not rely on an annotated score or audio representation, which can be expensive to obtain, severely limiting the breadth of audio recordings that such a method could be applied to.   Additionally, much previous comparison work has been limited to the comparison of a few pieces \citep{LiemHanjalic:2015} or a single instrument \citep{Sapp:2007, Sapp:2008, Liebman:2012}.  In contrast, HMDS allows for the analysis of systematic similarities of \textit{many} pieces of \textit{orchestral} recordings, and such large-scale analysis is critical to drawing musicological conclusions \citep{Cook:2005}.  Indeed, it would be prohibitive for a musicologist to listen to and compare pairwise 37 pieces recorded by 10 different orchestras, as we consider here.  The comparison of recordings in this work is on a much larger scale in terms of the number of pieces and the number of instruments (orchestral works) than the majority of previous work \citep{PeperkampEtAl:2017, LiemHanjalic:2015, Sapp:2007, Sapp:2008, Liebman:2012}.  Finally, HMDS is one of the few truly statistical, model-based approaches for orchestral recording comparison, as most existing approaches rely on purely algorithmic approaches \citep{PeperkampEtAl:2017, LiemHanjalic:2015, Liebman:2012}.    In particular, our approach in HMDS allows us to explore several of the findings and questions raised in \citet{Liebman:2012}, for example, if newer recordings are less idiosyncratic than older ones and the importance of geographic location in performance style, for a much larger, orchestral setting.  While we do not specifically perform hypothesis tests in this work, the model-based nature of HMDS allows for natural extensions to hypothesis tests on such musical questions of interest, which is a feature lacking in previous music information retrieval based approaches.  

The audio processing procedure and musical metrics are discussed in Section \ref{sec:audio}.  
In Section \ref{sec:model}, we first review BMDS and then develop our HMDS model and provide an algorithm for posterior approximation. 
In Section \ref{sec:results} we fit an HMDS model to the audio recordings of all 9 Beethoven symphonies recorded by 10 different orchestras.  We interpret the results and evaluate quantitatively the differences among the orchestras in the context of known, qualitative information. Our results recover 
several expected systematic similarities between orchestras, as well as identify some more novel results. For example, we find that modern recordings exhibit a high degree of similarity to each other, as compared to older recordings.
Conclusions and directions for future work  are discussed in Section \ref{sec:conc}.

\section{Audio Feature Extraction}\label{sec:audio}
\noindent  Fully parametric models of raw audio data can be highly complex, especially for our goal of identifying differences between orchestras. We thus propose to reduce the audio recordings for each piece to pairwise distances between orchestras, based on different musical characteristics of the recordings.
Specifically, for each recording we create three positive functions of time, representing tempo, volume dynamics, and timbre over the duration of the recording. For each of these three audio characteristics and for each piece, a distance is computed between each pair of orchestras using the Hellinger metric distance between the corresponding functions. In the remainder of this section, we motivate this proposed audio processing methodology, starting with identification of musical features of interest, then the processing and aligning of the audio recordings and finally comparison of the the audio features to form distance matrices.  

\subsection{Data and Features of Interest}

Our original data consist of audio recordings of all nine Beethoven symphonies recorded by 10 different orchestras (\autoref{pieces}) and the recordings span from the 1950s to 2016.  Each movement is treated as a separate piece, resulting in a total of 37 pieces.  For this work, we consider three main musical features of interest: tempo, dynamics and timbre.  Overall, we are interested in extracting features that represent artistic or expressive choices made by conductors and orchestras, rather than features that are specific to the recording process.  For example, the overall volume of a recording is a function of the microphone placement during the recording process, and is not an artistic choice made by the conductor.  For all of the musical features considered, we attempt to isolate and remove artifacts of the recording process to focus on expressive musical features.  

\begin{table}[h]
\centering 
   \caption{Orchestras, conductors and recordings years for the audio recordings of the 9 Beethoven symphonies considered in this work.}
\begin{tabular}{|c|c|c|}\hline
Orchestra & Conductor & Recording Years  \\\hline\hline
Academy of Ancient Music & Hogwood & 1986-1989   \\\hline
Berlin Philharmonic & Rattle & 2016  \\\hline
Berlin Philharmonic & von Karajan & 1982-1984  \\\hline
Chicago Symphony Orchestra & Solti & 1991  \\\hline
Leipzig Gewandhaus Orchestra & Masur & 1989-1993  \\\hline
London Symphony Orchestra & Haitink & 2006  \\\hline
NBC Symphony Orchestra & Toscanini & 1939-1952  \\\hline
New York Philharmonic & Bernstein & 1961-1967  \\\hline
Philadelphia Orchestra & Muti & 1988-2000   \\\hline
Vienna Philharmonic & Rattle & 2012 \\\hline
\end{tabular}
\label{pieces}
\end{table}%

Tempo is the speed at which a piece is performed and often varies over the course of an orchestral piece of music. We are interested in relative tempo changes between orchestras and not in the overall speed of a recording. For example, if the score for a given piece calls for the tempo to accelerate at a specific point in the piece, one orchestra may accelerate over only one measure of music, while another orchestra may accelerate over three measures of music.  Or, during the \textit{accelerando} denoted in the musical score, one orchestra may double their tempo, while another orchestra may barely increase their speed at all. These types of relative tempo changes between orchestras are examples of tempo features of interest.

Dynamics refer to the relative changes in volume of an orchestra over the course of a piece and we are interested in relative dynamic dissimilarities between orchestras, rather than the overall volume. For example, suppose the score for a given piece calls for a \textit{crescendo}, or increase in volume, at a specific point in the piece.  One orchestra may play twice as loud at the end of the crescendo as they did at the beginning of the crescendo, while another orchestra may not noticeably increase their volume at all over the crescendo.

Musical timbre refers to the quality or color of an orchestra's sound.  For example, a violin and a trombone have different timbres, and thus sound different from each other, even when playing the same note pitch.  Individual orchestra members contribute to the overall differences in timbre between different orchestras and for this work, we consider the global timbre of the entire orchestra as a feature of interest. While several methods for analyzing timbre exist \citep{Sueur:2018, Grachten:2013, Logan:2000}, we consider spectral flatness as a proxy for the timbre of the orchestra.  Spectral flatness is a measure of the tonality of a sound, where a spectral flatness of 1 means that there are equal amounts of energy spread throughout the entire spectrum (white noise), while a spectral flatness close to 0 indicates that the energy in the audio signal is concentrated in only a few frequency bands, approaching a pure tone \citep{seewave:2018}.  It is important to note that timbre is an inherently perceptual property of sound and can be difficult to define precisely, thus, spectral flatness is really a measure of one aspect that influences timbre, rather than a direct measure of timbre itself \citep{Muller:2015}. 

Based on our measure of spectral flatness, a difference in timbre between two orchestras corresponds to a difference in the spread of energy across the spectrum.  For example, at a specific point in a piece, the flutes in one orchestra may play as loud as the lower pitched instruments, resulting in a spread of energy across frequency bands and a high spectral flatness.  On the other hand, an orchestra where only lower instruments play at a loud volume for the same part in the piece would have spectral energy concentrated in fewer frequency bands and thus a lower spectral flatness. Spectral flatness can be sensitive to the recording technology used to produce the audio signal.  For example, older recordings that were converted from analog signals may have less energy in the upper frequencies, due to the audio conversion, and thus lower spectral flatness values compared to modern recordings that do not clip the upper frequencies.  However, we believe that spectral flatness is a good initial proxy for the relative timbre features of an orchestra.

\subsection{Audio Processing and Alignment}
\noindent In this subsection, we describe the specific audio processing steps to transform the original audio recordings into data representations that can be used to calculate pairwise distances for each of the musical features or metrics of interest described above.  Our procedure consists of three steps: spectral pre-processing, alignment and calculation of features, each of which will be described in detail below.


\subsubsection{Spectral Pre-Processing}
Before calculating the musical metrics of interest, we need a musically meaningful representation of the audio signals that will allow for the calculation of these features.  For example, we cannot determine the tempo of an orchestra for a given piece from the raw audio signal, so we will need another data representation that facilitates musical feature extraction.  We use two different representations of the audio signal, the spectrogram and the chromagram. The spectrogram represents the power spectrum of the audio signal over the entire frequency range and is found via the Short-Time Fourier Transform (STFT).  The spectrogram contains the energy distribution of the audio signal over time and is used to calculate the spectral flatness metric. The spectrogram for orchestra $i$ for piece $p$ is a $F$ x $T$ matrix, $S_{ip}(f,t)$, which represents the magnitude of the $f^{th}$ frequency band at time $t$. For all orchestras and all pieces, the frequency resolution of the spectrogram representation is 5 Hz and the temporal resolution is 0.1 seconds.  We only use the magnitude information from the spectrogram and ignore the phase information.

The chromagram representation \citep{Muller:2015} can be calculated from the spectrogram and is used to calculate the tempo and dynamics metrics. The chromagram aggregates the amplitude of each frequency bin in the spectrogram across octaves to give one amplitude for each pitch in the twelve tone scale.  This aggregation is robust to differences in instrument balance and intonation between orchestras.  The chromagram is a 12 x $T$ matrix, where each row corresponds to one note pitch, or pitch class.  For orchestra $i$ and piece $p$, let $\{f_q\}$ be the set of frequency bands that correspond to pitch class $q$.  For example, $q = 11$ corresponds to a $B^{\flat}$ note pitch, in any octave.  Then, the chromagram, $\bm \psi_{ip}$, can be calculated from the spectrogram as follows:
\begin{equation}
\psi_{ip}(q,t) = \sum_{k\in \{f_q\}} S_{ip}(k, t).
\end{equation}
\noindent The 12-dimensional chromagram representation at time $t$, $\psi_{ip}(:, t)$, can be  thought of as  approximating the notes that orchestra $i$ plays at time $t$. Example chromagrams are shown in \autoref{fig:chroma}.

In summary, we calculate the spectrogram, $\bm{S}_{ip}$, and chromagram, $\bm{\psi}_{ip}$, representations for all orchestras $i$ and pieces $p$.  All of the musical features of interest are derived from these two representations; the tempo and dynamics features are calculated from the chromagram representation and the timbre features are calculated from the spectrogram representation.

\subsubsection{Alignment}

\begin{figure}
\centering
\begin{subfigure}{.45\textwidth}
  \centering
  \includegraphics[width=\linewidth]{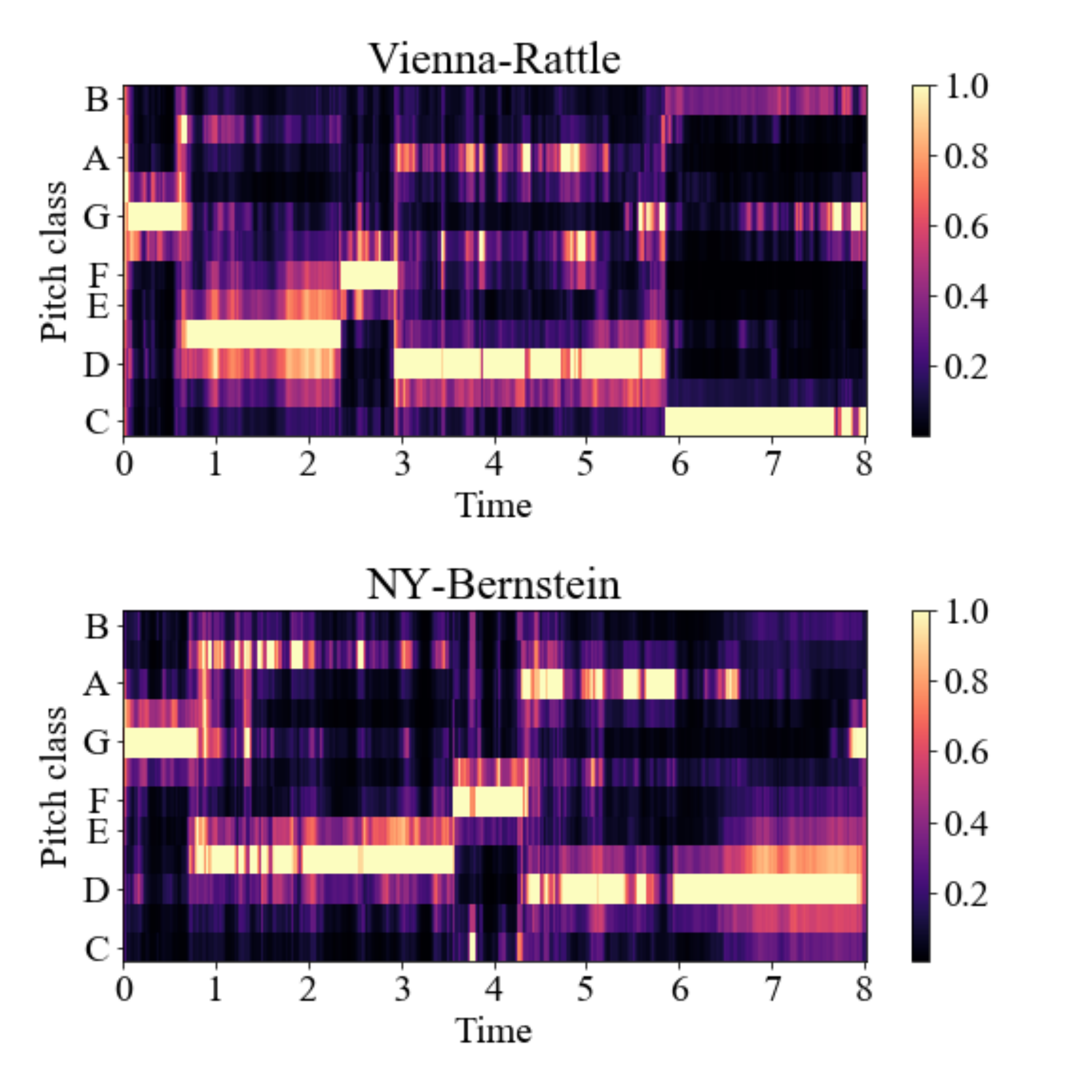}
  \caption{Unaligned chromagrams.}
  \label{fig:chrom-unaligned}
\end{subfigure}%
\begin{subfigure}{.45\textwidth}
  \centering
  \includegraphics[width=\linewidth]{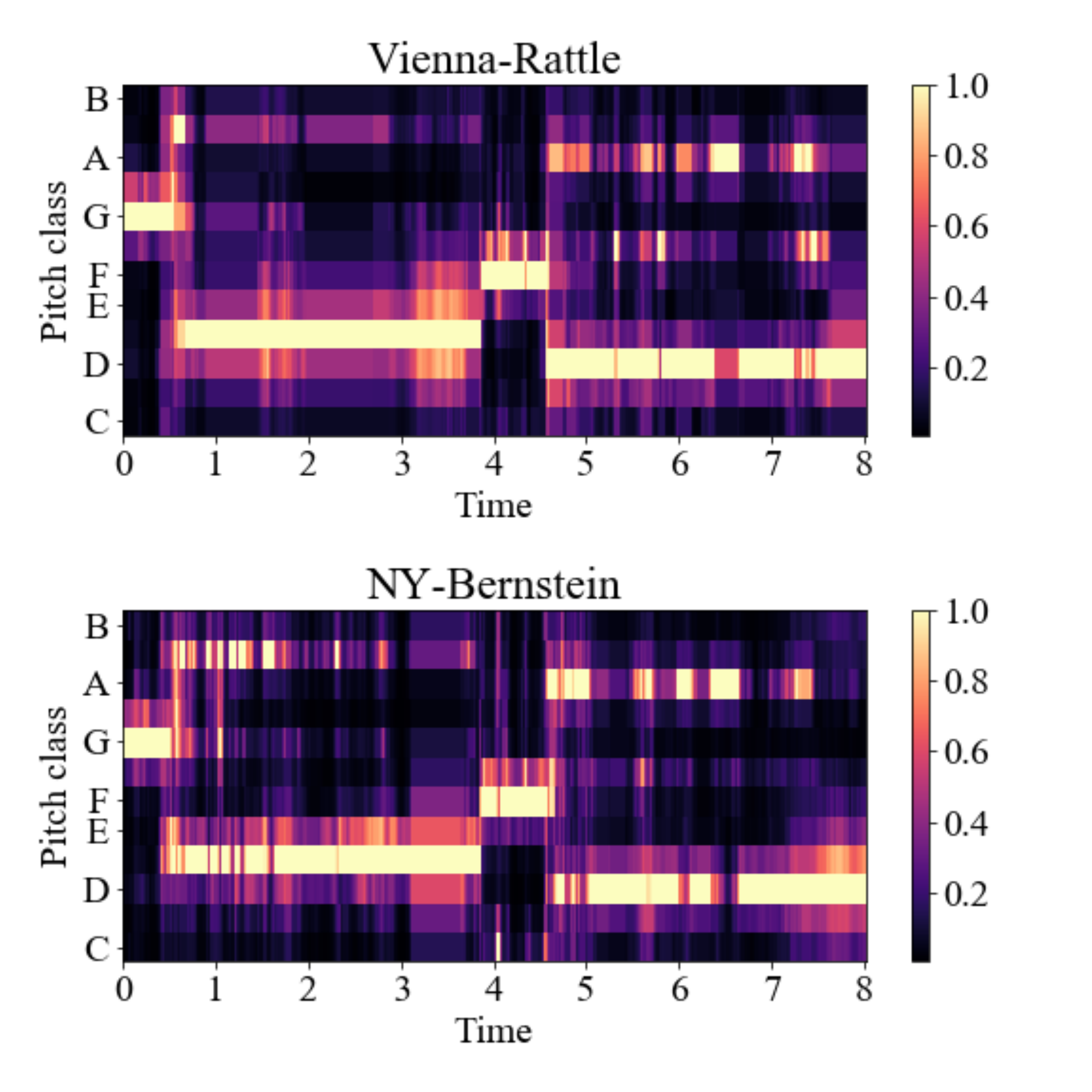}
  \caption{Aligned chromagrams.}
  \label{fig:chrom-aligned}
\end{subfigure}
\caption{(a) Unaligned and (b) aligned chromagrams  for the opening of  Beethoven No. 5 - Mvmt. 1 by the Vienna Philharmonic and the NY Philharmonic. The unaligned chromagrams show that the two orchestras play the same notes pitches, but at different speeds and for different durations, so these two orchestras do not perform the same part of the piece at the same point in time.  However, after alignment, the two orchestras do perform the same part of the piece at the same point in time and musical features can be extracted.}
\label{fig:chroma}
\end{figure}

For all pieces $p$, orchestra $i$ and orchestra $j$ play the same notes in the same order.  However, different orchestras may play these sequences of notes at different speeds.  For example, in \autoref{fig:chrom-unaligned}, the Vienna Philharmonic and the NY Philharmonic play the opening of Beethoven's Fifth Symphony at different tempos; Vienna only holds the opening $G$ pitch for about half a second and the following $E^{\flat}$ pitch from 0.5 to 2.5 seconds, while NY plays the opening $G$ for nearly an entire second and sustains the following $E^{\flat}$ pitch for nearly three seconds. Our goal is to compare differences in tempo, dynamics and timbre when each orchestra is performing the \textit{same part of each piece} to assess differences in artistic or expressive aspects of the performance.  Before calculating the musical features of interest, we then need to temporally align the spectrogram and chromagram representations so that on piece $p$, at time $t$ for orchestra $i$ and at time $t'$ for orchestra $j$, these two orchestras are performing the same part of the piece.  

Since the chroma vectors approximate the notes that an orchestra plays at time $t$, we want to find the warping path of indices, $w(t)$, such that $\psi_{ip}(:, t) = \psi_{jp}(:, w(t))$ for all $t$, subject to the constraints that $w(t)$ is monotonically increasing and that the orchestras start and end at the same point in the piece, that is, $\psi_{ip}(:, 0) = \psi_{jp}(:, 0)$ and $\psi_{ip}(:, T_{ip}) = \psi_{jp}(:, T_{jp})$, where $T_{ip}$ and $T_{jp}$ are the lengths of piece $p$ for orchestras $i$ and $j$, respectively. This problem can be solved via dynamic time warping. Dynamic time warping finds a non-linear warping path, $w(t)$, between the two chromagrams and is frequently used in music information retrieval \citep{Muller:2015, PeperkampEtAl:2017, ThornburgEtAl:2007, Ellis:2007, KirchhoffLerch:2011, KammerlEtAl:2014}. Dynamic time warping has also been used frequently within the larger context of MDS \citep{Kruskal:1983}.

Following  \cite{PeperkampEtAl:2017}, we align each piece to a reference MIDI recording from  \citet{Kunstderfuge} (MIDI is a symbolic music representation, and is simplified compared to the audio orchestral recordings). The result of the dynamic time warping is a warping path, $w_{ip}(t)$, for each orchestra $i$ for each piece $p$, relative to the reference MIDI recording. Then, we have that $\psi_{ip}(:, w_{ip}(t)) \approx \psi_{jp}(:, w_{jp}(t))$ and $S_{ip}(:, w_{ip}(t)) \approx S_{jp}(:, w_{jp}(t))$ for all $i$, $p$ and $t$. This equivalence is approximate, as specific chroma amplitudes can and do differ by orchestra; these differences correspond to variation in dynamics and timbre by orchestra, for example.  After the alignment, however, all orchestras perform the same part of the piece at the same time.  For simplicity, we also normalize the time by the length of each piece for each orchestra, so that $t\in [0,1]$ for all orchestras $i$ and pieces $p$. We can now use the aligned chromagrams and spectrograms to calculate our musical features of interest.

\subsubsection{Calculation of Audio Features}

\begin{figure}
\centering
\begin{subfigure}{.42\textwidth}
  \centering
  \includegraphics[width=\linewidth]{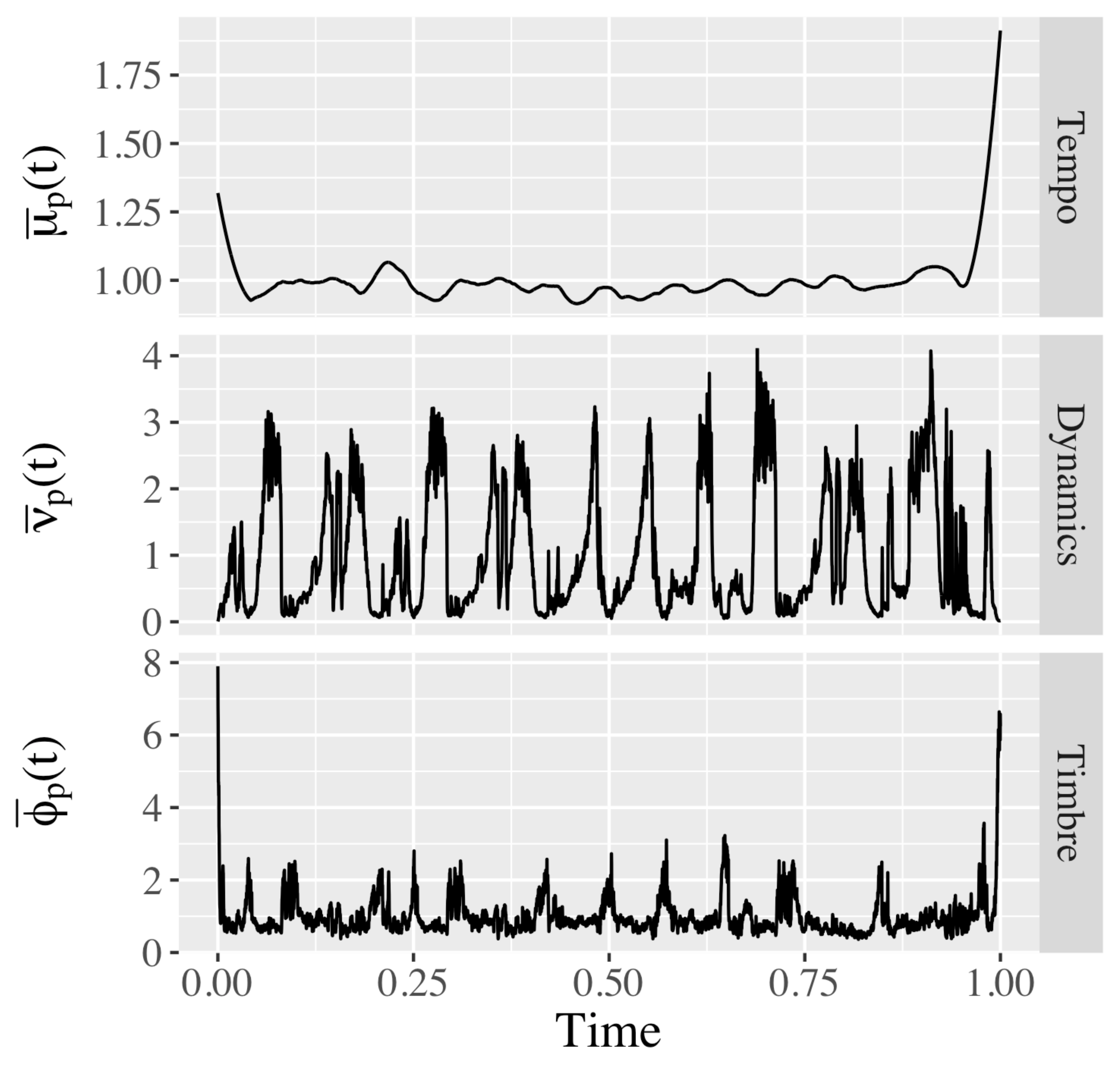}
  \caption{Mean metric functions.}
\end{subfigure}%
\begin{subfigure}{.5\textwidth}
  \centering
  \includegraphics[width=\linewidth]{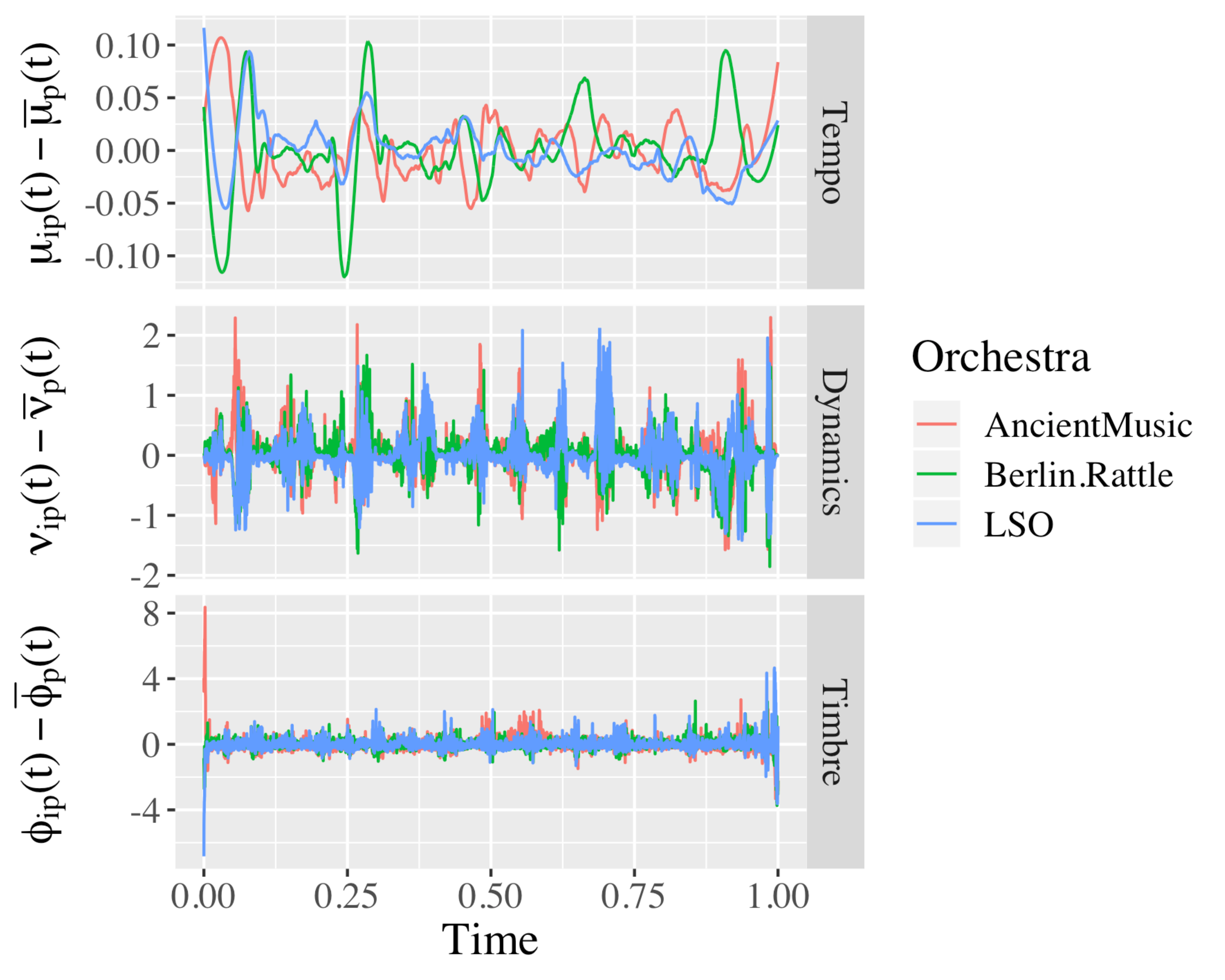}
  \caption{Deviations from mean metric functions.}
\end{subfigure}
\caption{(a) Mean metric functions and (b) deviations from mean metric functions by orchestra, for Beethoven Symphony No. 6, Mvmt. 1.  The time is normalized to be between 0 and 1, since all recordings are already aligned to the same reference recording.}
\label{fig:curves}
\end{figure}

After the alignment of our data representations, we are ready to calculate the specific musical metrics of interest.  Starting with the aligned spectrograms and chromagrams, we calculate the tempo, dynamics and timbre features as normalized functions that are used to calculate pairwise distances between orchestras. Let $\tilde{\bm \psi}_{ip}$ denote the aligned chromagrams, that is, $\tilde{\psi}_{ip}(:, t') = \psi_{ip}(:, w(t))$, and similarly for the aligned spectrograms, $\tilde{\bm S}_{ip}$, that will be used for the calculation of these musical metric functions.  Example functions are shown in \autoref{fig:curves}.

The tempo function can be calculated using the warping path from the dynamic time warping alignment. That is, the tempo function for orchestra $i$ on piece $p$ is 
$$\mu_{ip}(t) \propto w'_{ip}(t) = \dfrac{d w_{ip}(t)}{d t},$$
\noindent where $\mu_{ip}(t)$ is the ratio of the tempo of orchestra $i$ on piece $p$ at time $t$ relative to the reference recording for piece $p$.  This means that at time $t$ for piece $p$, if $\mu_{ip}(t) = 4$ and $\mu_{jp}(t) = 2$, then orchestra $i$ is playing two times as fast as orchestra $j$. Likewise, $\mu_{ip}(t)  < 1$ means that orchestra $i$ is performing piece $p$ slower than the reference recording at time $t$. The tempo curve is normalized such that
$$\mu_{ip}(t) = \dfrac{w'_{ip}(t) }{\sum_{s= 0}^1 w'_{ip}(s) }.$$

The dynamics function can be calculated using the sum of the magnitudes of the aligned chromagram at each point in time, divided by the average volume for the entire piece.  That is, the dynamics are calculated as
$$v_{ip}(t) \propto \dfrac{\sum_{q=1}^{12}\tilde{\psi}_{ip}(q, t)}{\frac{1}{12}\sum_{q=1}^{12}\sum_{s=0}^1\tilde{\psi}_{ip}(q,s)}.$$
\noindent We normalize the dynamics curves by the average volume for that orchestra for piece $p$, as we are not interested in the overall volume of each orchestra.  Again, we define $v_{ip}(t)$ to be a normalized function, so that $\sum_{s=0}^1 v_{ip}(s) = 1.$ Then, $v_{ip}(t) > v_{jp}(t)$ means that relative to each orchestra's respective average dynamic for piece $p$, orchestra $i$ is playing louder than orchestra $j$ at time $t$. 

Spectral flatness is the ratio of the geometric mean to the arithmetic mean and can be calculated as
$$\phi_{ip}(t) \propto F\times \dfrac{\sqrt[F]{\prod_{f=1}^F \tilde{S}_{ip}(f, t)}}{\sum_{f=1}^F \tilde{S}_{ip}(f, t)},$$

\noindent where $\tilde{S}_{ip}(f, t)$ is the relative amplitude of the  $f^{th}$ frequency of the aligned spectrogram and $F$ is the total number of frequencies for the aligned spectrogram \citep{seewave:2018} and we have that $\sum_{s=0}^1 \phi_{ip}(s) = 1.$  For the spectral flatness features, $\phi_{ip}(t) > \phi_{jp}(t)$ means that the energy of orchestra $j$ is concentrated in a smaller number of frequency bands than orchestra $i$ for piece $p$ at time $t$.    


In summary, after transforming the audio signals to spectrogram and chromagram representations, temporally aligning the representations and calculating the musical metrics for tempo, dynamics and timbre, we have musical feature functions for tempo, $\mu_{ip}(t)$, dynamics, $v_{ip}(t)$ and timbre, $\phi_{ip}(t)$, for all orchestras $i = 1, \ldots, 10$, pieces $p = 1, \ldots, 37$ and time $t\in[0,1]$.  These functions are normalized to sum to 1, such that $\sum_{s = 0}^1 \mu_{ip}(s) = \sum_{s=0}^1 v_{ip}(s) = \sum_{s=0}^1 \phi_{ip}(s) = 1.$

\subsection{Comparison of Audio Features}
Now that we have normalized functions for each orchestra, for each piece and for each musical feature, we can calculate pairwise distance matrices using a density-based distance measure.  We calculate the pairwise Hellinger distance between all orchestras for each piece to obtain $M$ pairwise distance matrices for each musical metric. The Hellinger distance is a commonly used density-based distance measure and for discrete distributions $P = (p_1, \ldots, p_K)$ and $Q = (q_1, \ldots, q_K)$ can be calculated as

\begin{equation}
H(P, Q) = \dfrac{1}{\sqrt{2}} \sqrt{\sum_{k=1}^K \left(\sqrt{p_k} - \sqrt{q_k}\right)^2}. \\
\end{equation}

\noindent After calculating the pairwise Hellinger distance between orchestras, the distances within each metric are normalized to be between 0 and 1 over all pieces. The end result of the audio feature extraction procedure is $M$ distance matrices for each of the three musical metrics, tempo, $D^{\mu}$, dynamics, $D^{v}$, and timbre, $D^{\phi}$,  where each distance matrix is of dimension $N\times N$, where $N =10$ is the number of orchestras and $M=37$ is the number of pieces.  Then, for a specific musical metric, $y_{ijp}$ is the Hellinger distance between orchestra $i$ and orchestra $j$ on piece $p$ for that musical metric.  We treat these $y_{ijp}$ distances as our observed data for the modeling described in the next section.  Additionally, each individual distance matrix is symmetric, so $y_{ijp} = y_{jip}$, for all orchestras $i$ and $j$, and pieces $p$.

The overall audio processing methodology can thus be summarized as follows. We start with audio recordings of all nine Beethoven symphonies by $N$ different orchestras. Each movement of a symphony is treated as a separate piece, resulting in $M$ pieces.  Using spectrogram and chromagram representations of the audio signals, we temporally align the orchestra representations by piece using dynamic time warping, and calculate tempo, dynamics and timbre functions by piece. Then, we use the Hellinger density-based distance measure to calculate $M$ pairwise replicate distance matrices (each of dimension $N\times N$) for each of the three musical metrics.  The data and accompanying code is available in the Supplementary Material \citep{HMDS:supp} and online at \url{https://github.com/aky4wn/HMDS}.

\section{Hierarchical Multidimensional Scaling}\label{sec:model}

In this section, we develop Hierarchical Multidimensional Scaling, a statistical model for a sample of pairwise distance matrices among a common set of objects.  The purpose of the model is to identify patterns in the distance matrices that are consistent throughout the sample as well as to quantify the variation of the distance matrices around these patterns.  For example, in our application, each matrix in our sample represents the pairwise distance between orchestras for a given piece for a specific musical metric, either tempo, dynamics or timbre. Our model-based approach is a modification and extension of the approach of \citet*{Oh-Raftery:2001}, who developed a probability model for a single dissimilarity matrix.

\subsection{Bayesian Multidimensional Scaling}

\citet{Oh-Raftery:2001} propose Bayesian Multidimensional Scaling (BMDS), a model-based version of classical MDS.  For a \textit{single} data matrix of pairwise distances between objects, the goals of BMDS are to find a low-dimensional representation of the objects of interest in Euclidean space and to measure the discrepancy between the Euclidean space and the observed distances.  BMDS assumes that observed pairwise distance measurements are equal to a true distance measure plus observational noise, where the true distance measure is the Euclidean distance between latent embedding vectors for the pair.  Let $N$ be the number of objects or entities of interest and $y_{ij}$ be the observed distance between object $i$ and object $j$. Then, the BMDS model is defined as follows: 
\begin{equation}\label{eq:BMDS}
\begin{split}
y_{ij} & \sim \mbox{TruncNorm}(||X_i - X_j||_2, \; \sigma^2), \enspace j > i, \; i,j = 1, \ldots, N, \\
\end{split}
\end{equation}

\noindent independently, where $X_1, \ldots, X_N$ are unobserved latent vectors in $r$-dimensional Euclidean space, one for each object $i$, and $\sigma^2$ is an unknown scale parameter. $\mbox{TruncNorm}$ is the normal density, truncated to be above 0.  Note that for BMDS, the error term $\sigma^2$ represents the deviation of the observed distances from Euclidean distances, which could be attributed to either measurement error or misspecification of the (true) distances being Euclidean.

\citet{Oh-Raftery:2001} describe a Markov chain Monte Carlo algorithm for approximating the posterior distribution of $X_1, \ldots, X_N$ and $\sigma^2$, conditional on the observed distance data $\{y_{ij} : \; j > i\}$. They specify independent priors for each of the unknown parameters. The latent $X_i$ vector for each object is assumed to come from an independent, $r$-dimensional normal distribution with diagonal covariance matrix $\Lambda$, where the diagonal elements of $\Lambda$ are inverse-gamma distributed.  The error term, $\sigma^2$, is assigned an inverse-gamma prior distribution for conjugacy.  The original BMDS model does not consider replications or multiple distance matrices, though later extensions do for a specific preference data application \citep{Park:2008}, as discussed in Section \ref{sec:intro}.

\subsection{Hierarchical MDS for Multiple Distance Matrices}\label{sec:HMDS}

Given only a single distance matrix of observed distances, the BMDS model cannot distinguish between measurement error and the degree to which the systematic distances between objects differ from being Euclidean. For example, a large estimated $\sigma^2$ value in \autoref{eq:BMDS} could indicate a large amount of measurement error in the observed pairwise distances \textit{or} that the observed distances are not well represented by Euclidean distances between latent vectors (or some combination of these two factors).  With \textit{replicate distance matrices}, however, there is sufficient information to distinguish a non-Euclidean mean distance from across-sample measurement variation. We quantify the variation of the replicate distance matrices around a common mean distance matrix with the following hierarchical MDS (HMDS) model.  Let $y_{ijp}$ be the observed pairwise distance between entity $i$ and entity $j$ for observation $p$, where there are $N$ total entities and $M$ total replicate distance matrices.   The HMDS model is given in \autoref{eq:HMDS}:
\begin{equation}\label{eq:HMDS}
\begin{split}
y_{ijp} & \sim \mbox{Gamma}\left(\psi, \; \dfrac{\psi}{\tau_p\delta_{ij}}\right), \enspace j > i, \; i,j = 1, \ldots, N, \; p = 1, \ldots, M, \\
\end{split}
\end{equation}

\noindent independently across pairs and replicates, where $\psi, \tau_1, \ldots, \tau_M$ and $\{\delta_{ij} : j > i\}$ are parameters to be estimated. The gamma distribution is parameterized such that the mean of $y_{ijp}$ is $\tau_p\delta_{ij}$ and the variance of $y_{ijp}$ is $(\tau_p^2\delta_{ij}^2)/\psi$. 
Each $\tau_p$ is a scale parameter for replicate distance matrix $p$ that allows for each matrix to have a different ``potential'' for variation.  The $\{\delta_{ij} : j > i\}$ parameters represent the systematic dissimilarity between entities $i$ and $j$ across all $M$ replicate matrices, while the $\psi$ parameter serves as an overall scale factor.  Note that $y_{ijp} = y_{jip}$ for all entities $i$ and $j$ and replicate distance matrices $p$, so we only model explicitly the unique entries in each distance matrix.

Inclusion of the $\tau_1, \ldots, \tau_M$ parameters is important in our application,  as we expect some pieces to have more inherent opportunities for variation than other pieces.  For example, some pieces have numerous vague tempo markings that allow for a good deal of artistic interpretation and tempo variation between orchestras, as compared to other pieces that do not have many denoted changes in tempo. However, a piece's ``potential'' for variation is a characteristic of the piece and is separate from the systematic differences between orchestras, and must be handled accordingly. This potential for variation scales both the mean and the variance of the observed distances.

The proposed HMDS model differs in several ways from the BMDS model. First, the gamma distribution for the observed pairwise distances is a more natural choice for a positive random quantity than the truncated normal distribution of BMDS.  The gamma distribution allows for skew in the observed pairwise dissimilarities and facilitates straightforward parameter estimation and inference, as will be described below.  Second, the HMDS model in \autoref{eq:HMDS} does not restrict the $\{\delta_{ij}: j > i\}$ parameters to correspond to Euclidean distances.  This relaxes the assumptions of BMDS and is important in many applications, including our comparison of orchestral recordings, where the distance metrics used to compare objects are known to be non-Euclidean.

The HMDS parameters can be estimated with maximum likelihood estimation.  The MLE estimates for the $\delta_{ij}$ and $\tau_p$ parameters satisfy the following system of equations: 
$$\hat{\delta}_{ij}^{MLE} = \dfrac{1}{M}\sum_{p=1}^M \dfrac{y_{ijp}}{\hat{\tau}_p^{MLE}}, \enspace \hat{\tau}_p^{MLE} = \dfrac{2}{N(N-1)}\sum_{i = 1}^N\sum_{j > i}\dfrac{y_{ijp}}{\hat{\delta}_{ij}^{MLE}}.$$

\noindent The MLE for $\psi$ can be found by iteratively solving the following equation:
$$\dfrac{\Gamma'(\psi)}{\Gamma(\psi)} - \log \psi = \dfrac{2}{N(N-1)M}\left[1 + \sum_{i=1}^N\sum_{j > i}\sum_{p=1}^M \left(\log\left(\dfrac{y_{ijp}}{\tau_p\delta_{ij}}\right) - \dfrac{y_{ijp}}{\tau_p\delta_{ij}}\right)\right],$$

\noindent where $\Gamma'(\psi)$ is the digamma function, $\Gamma'(\psi) = \tfrac{d \Gamma(\psi)}{d\psi}$. Importantly, note that the MLE estimates, $\hat{\delta}_{ij}^{MLE}$, might not be distances.

We choose to perform Bayesian inference for parameter estimation.  Bayesian inference in the HMDS model naturally allows for parameter uncertainty estimates.  Additionally, the space of $N$ x $N$ distance matrices is quite high-dimensional and complex, and Bayesian inference provides shrinkage towards a lower dimensional space.  Finally, the choice of a hierarchical model for the $\{\delta_{ij} : j > i\}$ parameters centered around a Euclidean space can aid in parameter interpretation.  That is, we can specify a prior that puts the $\{\delta_{ij} : j > i\}$ parameters near some actual distances that satisfy the triangle inequality.  To that end, we model the $\{\delta_{ij} : j > i\}$ parameters as
\begin{equation}\label{eq:delta-HMDS}
\delta_{ij} \sim \mbox{Inv-Gamma}\left(\gamma, \;  (\gamma + 1) ||X_i - X_j||_2\right), \enspace j > i, \; i,j = 1, \ldots, N,  \\
\end{equation}
\noindent independently across pairs. The inverse-gamma distribution is parameterized such that the prior mode for $\delta_{ij}$ is $||X_i - X_j||_2$. The goal of the prior is to shrink the dissimilarity between orchestras $i$ and $j$ towards a distance metric that follows the triangle inequality.  The $||X_i - X_j||_2$ term fixes this metric space as an $N-1$ dimensional Euclidean space, though the $\gamma$ parameter allows for potentially substantial variation from this Euclidean distance. Notably, this variation about a Euclidean distance, represented by the $\gamma$ parameter in HMDS, is separate from the across-replicate sampling variability, represented by the parameter $\psi$.  This separation of sources of variation is in contrast to the BMDS model. 

This particular choice of an inverse-gamma prior further facilitates computation and interpretation.  The effect of the prior on estimation of the $\delta_{ij}$'s can be understood from the form of their full conditional distributions.  The conditional density of $\delta_{ij}$ given all other model parameters and the observed $y_{ijp}$ pairwise distances is
$$\delta_{ij}\Big | \psi, \gamma, \{\tau_p\}_{1}^M, \{X_i\}_{1}^N, Y \;  \sim \mbox{Inv-Gamma}\left(M\psi + \gamma, \; (\gamma + 1)||X_i - X_j||_2 + \psi\sum_{p = 1}^M\dfrac{y_{ijp}}{\tau_p}\right),$$
\noindent where $Y = \{y_{ijp} : i = 1, \ldots, N, j > i, p = 1, \ldots, M\}$. The mode of this conditional distribution is
$$\dfrac{\gamma + 1}{M\psi + \gamma + 1}||X_i - X_j||_2 \; + \; \dfrac{M\psi}{M\psi + \gamma + 1}\hat{\delta}_{ij}^{MLE},$$
\noindent so the $\delta_{ij}$ prior specified in \autoref{eq:delta-HMDS} shrinks the MLE estimate towards a set of Euclidean distances.

The full conditional distribution of the $\tau_p$ parameters, given all other model parameters and the observed $y_{ijp}$ pairwise distances is 
$$\tau_p\Big | \psi, \gamma, \{\delta_{ij} : j > i\}, \{X_i\}_1^N, Y \; \sim  \mbox{Inv-Gamma}\left(\dfrac{N(N-1)}{2}\psi + 1, \; \psi\sum_{i = 1}^N\sum_{j > i}\dfrac{y_{ijp}}{\delta_{ij}}\right),$$
\noindent and the conditional mode is
$$\dfrac{\psi N(N-1)}{\psi N (N-1) + 4}\hat{\tau}_{p}^{MLE}.$$
\noindent The conditional mode of $\tau_p$ depends on how close the observed pairwise distances (the $y_{ijp}$'s) are to the systematic dissimilarities (the $\delta_{ij}$'s) across all pairs, $j > i$.  That is, if $y_{ijp} = \delta_{ij} \; \forall j > i, i = 1, \ldots, N$, then $\hat{\tau}_p^{MLE} = 1$ and the variation of the $y_{ijp}$ distances is only scaled by $\psi$. In this case,  the observed distances are equal to the systematic dissimilarities and there is no ``potential'' for variation for replicate matrix $p$.  However, when $y_{ijp} > \delta_{ij}$ across object pairs for a given replication matrix $p$, then $\hat{\tau}^{MLE}_p > 1$ and that specific replication $p$ has a high ``potential'' for variation.

 For a full Bayesian analysis, we also need to specify priors for the remaining unknown parameters: 
\begin{equation}\label{HMDS-priors}
\begin{split}
X_{1}, \ldots, X_N &\overset{indep.}{\sim} \mathcal{N}_{N-1}(0, \Lambda) \\
 \psi & \sim \mbox{Gamma}(a_1,b_1) \\
 \gamma & \sim\mbox{Gamma}(a_2, b_2) \\
 \tau_1, \ldots, \tau_M &\overset{indep.}{\sim} \mbox{Inverse-Gamma}(\alpha, \beta) \\
\end{split}
\end{equation}

\noindent where $\Lambda$ is a diagonal matrix and $a_1, b_1, a_2, b_2, \alpha$ and $\beta$ are positive scalars.  We set  $a_1, b_1, a_2, b_2$ to 0.01, and $\alpha$ and $\beta$ to 1 in our application.  These choices of hyper-priors specify weakly-informative prior distributions for $\psi$, $\gamma$ and the $\tau_p$'s, following \citet{Gelman:2006}.  In particular, the choice of the shape parameter $\alpha$ less than or equal to 1 in the prior for the $\tau_p$ parameters means that we believe the main mass of the prior for each $\tau_p$ is pushed towards 0, favoring small values of $\tau_p$, while still maintaining heavy tails, allowing for potentially very large $\tau_p$ values.  As we expect there to be different potentials for variation by piece $p$, this inverse-gamma prior specification allows for a wide spread of possible $\tau_p$ values. We  use an empirical Bayes approach to setting $\Lambda$ by estimating the variance of embedding vectors from classical MDS performed for each piece.  A full sensitivity analysis of the hyper-parameter choices for the gamma and inverse-gamma priors is included in Section 2 of the Supplementary Material \citep{HMDS:supp}; we find that the overall results are not very sensitive to the choice of prior hyper-parameter values.  Similarly to BMDS, the $X_i$ vectors are assumed to explicitly follow a prior distribution here, which is in contrast to the usual point estimates for the embedding vectors that result from classical MDS.

The joint posterior distribution of model parameters can be approximated with a Markov chain Monte Carlo algorithm.  One such algorithm proceeds as follows:
\begin{enumerate}
\item For each $i = 1, \ldots, N, \; j > i$, simulate
$$\delta_{ij} \sim \mbox{Inv-Gamma}\left(M\psi + \gamma, \; (\gamma + 1) ||X_i - X_j||_2 + \psi\sum_{p = 1}^M \dfrac{y_{jip}}{\tau_p}\right).$$
\item For each $p = 1, \ldots, M$, simulate
$$\tau_p \sim \mbox{Inv-Gamma}\left(\alpha + \dfrac{N(N-1)}{2}\psi, \; \beta + \psi \sum_{i=1}^N\sum_{j > i}\dfrac{y_{ijp}}{\delta_{ij}}\right).$$
\item The remaining parameters, $\Big\{ X_1, \ldots, X_N, \; \psi, \; \gamma \Big\}$, can be updated via Metropolis-Hastings steps.  
\end{enumerate}

\noindent Full details of the Markov chain Monte Carlo algorithm are given in Section 1 of the Supplementary Material \citep{HMDS:supp}.

\subsection{Comparison to Classical MDS Approaches}

In the classical MDS literature, related approaches for 3-way data include INDSCAL (Individual Differences in Scaling) and IDIOSCAL (Individual Differences in Orientation Scaling) \citep{Carroll:1970}.  Both of these approaches extend classical MDS to repeated observations or measurements between entities.  However, unlike HMDS, a primary purpose of these methods is to identify a common low-dimensional subspace in which these distance matrices can be embedded, and to describe across-matrix heterogeneity in this subspace.  INDSCAL and IDIOSCAL both assume that there exists a group embedding space, $Z$, which represents commonalities among all entities.  Each entity then has a individual space, $X_i$, that is a transformation of the group space $Z$, that is $X_i = ZW_i$, where $W_i$ is a diagonal matrix with positive entries for INDSCAL and a general matrix for IDIOSCAL \citep{Borg:2018}.  Both methods iteratively update $Z$ and $W_i$ to minimize the least-squares difference between the observed and fit distances.  The weight matrix $W_i$ can be interpreted as the transformations that stretch and compress the group space $Z$ into each individual space $X_i$ \citep{Borg:2018}.  Similar to HMDS, this $W_i$ matrix in INDSCAL and IDIOSCAL allows for variation by replication.  INDSCAL and IDIOSCAL also share information across replicates by assuming a common group subspace, $Z$.

While these methods could be applied to our audio recording data, the objectives and assumptions of INDSCAL and IDIOSCAL are different from those of HMDS.  In particular, our primary goal with HMDS is to estimate a distance matrix (the $\delta_{ij}$'s) that represents consistent differences between pairs of orchestras across pieces, rather than to find a common, reduced-dimension subspace in which to represent the across-matrix variations.  Indeed, the $\delta_{ij}$'s of interest in HMDS need not be embeddable in a lower-dimensional Euclidean space.  Although each $X_i$ shares a common prior specification, the $X_i$ vectors are not constrained to lie in a common ``group space'' like in INDSCAL and IDIOSCAL.  The HMDS method is additionally applicable to distance matrices that are not Euclidean, which is especially important in our application, where the distances of interest are known to be non-Euclidean.  Finally, a primary, important difference between HMDS and classical MDS approaches is that our approach is model-based, which allows for a description of uncertainty in the parameter estimates, as well as information-sharing across the data matrices via the hierarchical model.  We thus view HMDS as an extension of BMDS to replicate distance matrices that focuses on different modeling goals than the classical INDSCAL and IDIOSCAL approaches.

\section{Analysis of Orchestral Distance Data}\label{sec:results}

In this section, we fit the HMDS model to the replicate orchestral distance matrices to explore systematic differences between orchestras across pieces.  We fit the HMDS model separately for each of the three musical distance metrics: tempo, dynamics and timbre.  We check the MCMC approximation outlined in Section \ref{sec:HMDS} and examine the goodness-of-fit of HMDS for our orchestral audio data.  Finally, we analyze the learned parameters for each musical metric and find that the HMDS model is able to recover musically expected systematic differences between orchestras across pieces, as well as suggest some unexpected similarities between orchestras.

\subsection{MCMC Approximation and Goodness-of-Fit}

The posterior distribution of parameters in the HMDS model can be approximated using a Markov chain Monte Carlo (MCMC) algorithm, such as the one described in Section \ref{sec:HMDS} or a more general MCMC algorithm \citep*{rstan}.  We fit the HMDS model in Rstan with the default No-U-Turn Hamiltonian Monte Carlo sampler.   Parameter values were randomly initialized and the chain was run for 30000 iterations, where the first 15000 iterations were discarded as burn-in and no thinning was performed.  The final 15000 iterations were retained for posterior inference.  We diagnose the posterior approximation with trace plots and effective sample size (ESS) diagnostics \citep{Stan-ref}.  The ESS diagnostic uses the estimated autocorrelation of the Markov chain to assess the Monte Carlo approximation error. Specifically, an ESS of $n_{eff}$ means that the
Monte Carlo error is approximately equivalent to what it would be from $n_{eff}$ independent simulations from the posterior distribution \citep{BDA}. 

Trace plots for a subset of the $\tau_p$ parameters for the tempo metric are shown in \autoref{fig:trace-tau-tempo} and trace plots for the remaining metrics are included in Section 2 of the Supplementary Material \citep{HMDS:supp}.  The trace plots are used to identify any evidence of non-stationarity of the Markov chain, which would indicate that the Markov chain would need to be run longer.  All trace plots appear stationary for the 15000 posterior simulations used for analysis.   The median ESS values across all $M$ pieces for the $\tau_p$ parameters are 407 for the tempo metric, 189 for the dynamics metric and 274 for the timbre metric.  While the ESS values for the $\tau_p$ parameters do indicate that the posterior Markov chains are mixing slowly, the trace plots across parameters and musical metrics appear stationary. 

\begin{figure}[htbp]
\begin{center}
\includegraphics[width =0.95\textwidth]{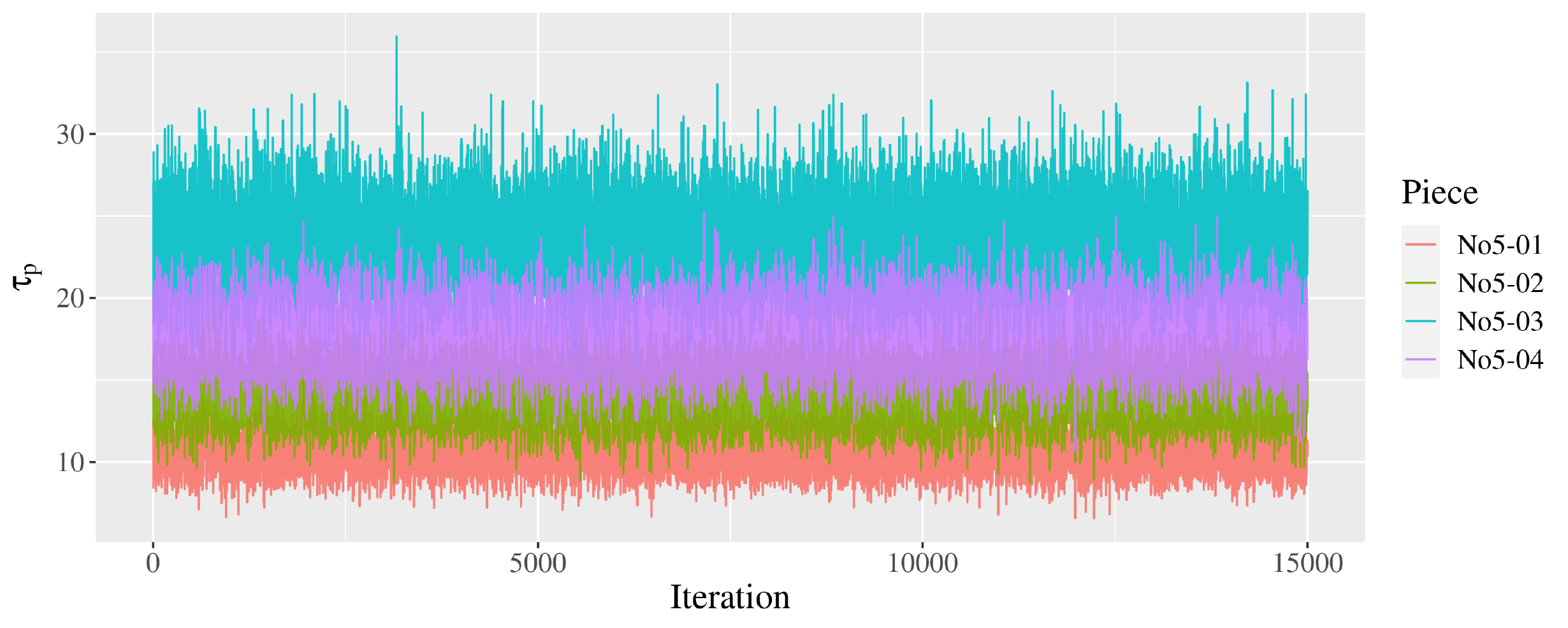}
\caption{Trace plot for Symphony No. 5 for $\tau_p$ for the tempo metric.}
\label{fig:trace-tau-tempo}
\end{center}
\end{figure}

\begin{figure}[htbp]
\begin{center}
\includegraphics[width = 0.95\textwidth]{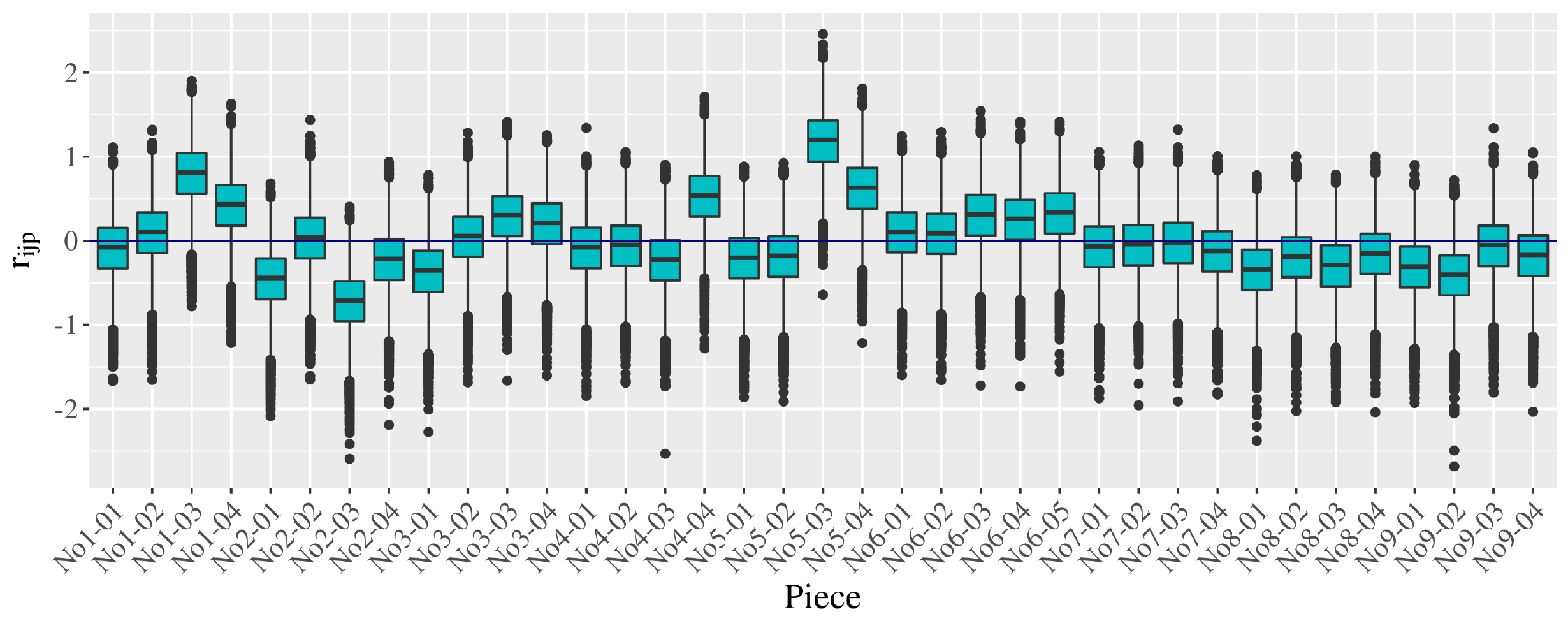}
\caption{Posterior predictive checks for the HMDS sampling model for one orchestra pair (Academy of Ancient Music and Vienna-Rattle) across all pieces for the tempo metric, $r_{ijp} = \log(\tilde{y}_{ijp}/ y_{ijp})$.}
\label{fig:tempo-y1-10}
\end{center}
\end{figure}

We also analyze the goodness-of-fit of the HMDS model for our orchestral audio data using posterior predictive checks \citep{BDA}.  We can evaluate the goodness of fit of the pair-specific sampling model by simulating posterior predictive values, $\tilde{y}_{ijp} \sim \mbox{Gamma}(\psi, \psi/(\tau_p\delta_{ij}))$, at each iteration of the Markov chain.  These values can be compared to the observed $y_{ijp}$'s to evaluate the fit of the pair-specific sampling model by computing $r_{ijp} =    \log (\tilde{y}_{ijp} /y_{ijp})$ for each simulated $\tilde{y}_{ijp}$, where the ratio accounts for differences in scale due to $\tau_p$.  The distribution of these ratios for one orchestra pair and all pieces for the tempo metric is displayed in \autoref{fig:tempo-y1-10}.  Additionally, we can evaluate the coverage of these ratio distributions.  For the tempo metric, 95.14\% of the 95\% highest-posterior density (HPD) intervals for each $r_{ijp}$ contain 0.  Likewise, 95.79\% and 95.38\% of the 95\% HPD intervals for $r_{ijp}$ contain 0, for dynamics and timbre, respectively. Overall, these results indicate that the sampling model for HMDS fits the data well.

\begin{figure}
\begin{center}
\includegraphics[width = 0.9\textwidth]{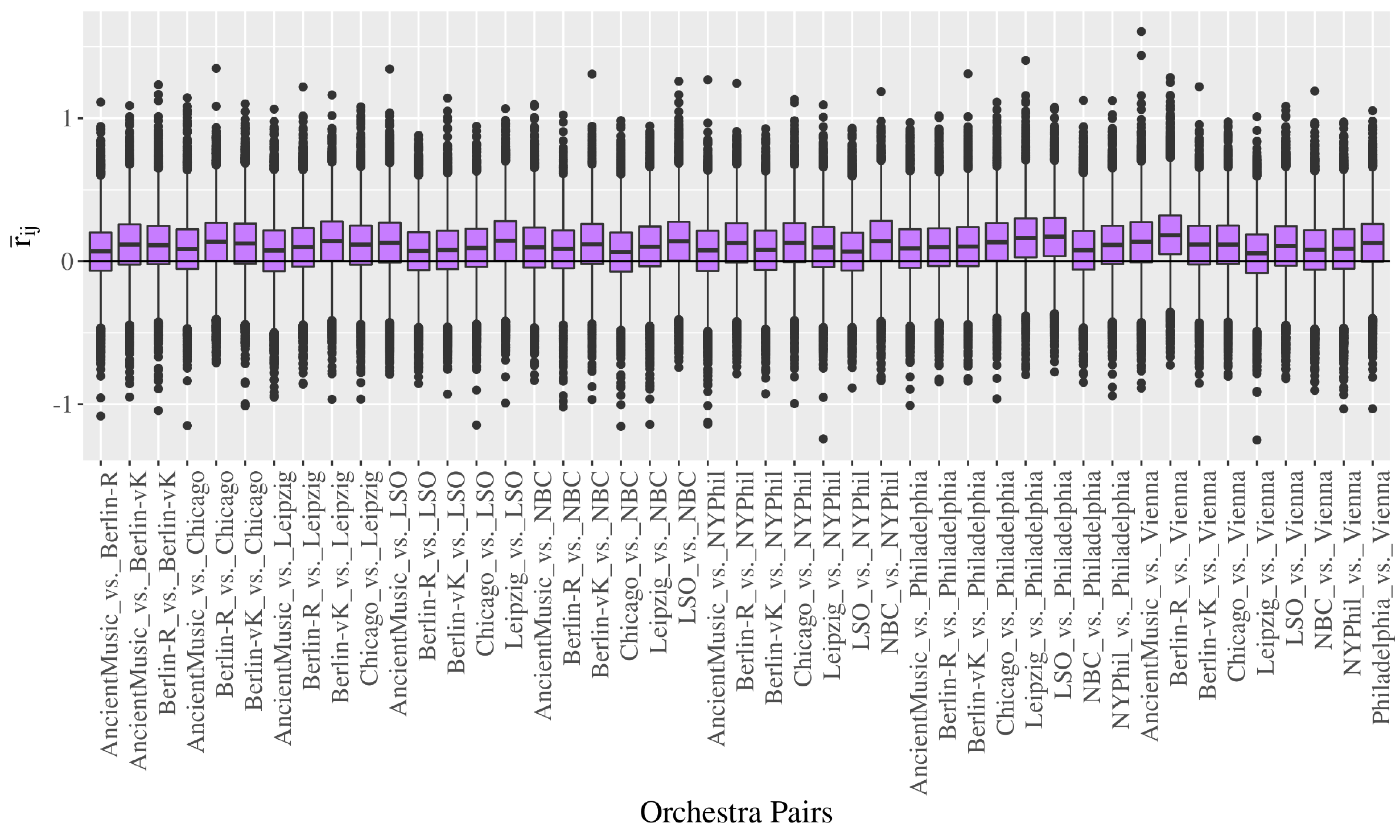}
\caption{Posterior predictive checks for the hierarchical sampling model for the tempo metric, averaged across all pieces.}
\label{fig:tempo-hier}
\end{center}
\end{figure}

Likewise, we can evaluate the goodness-of-fit of the hierarchical portion of HMDS with posterior predictive checks.  At each iteration of the Markov chain, we can calculate $||X_i - X_j||_2$, simulate $\tilde{\delta}_{ij} \sim \mbox{Inv-Gamma}(\gamma, \; (\gamma + 1) ||X_i - X_j||_2)$ and then simulate posterior predictive values, $\tilde{y}_{ijp} \sim \mbox{Gamma}(\psi, \psi/(\tau_p\tilde{\delta}_{ij}))$.  We compute $r_{ijp} =    \log (\tilde{y}_{ijp} /y_{ijp})$ for each simulated $\tilde{y}_{ijp}$ and the distribution of these ratios for all orchestra pairs, averaged across pieces (i.e.  $\bar{r}_{ij}$) is displayed in 
\autoref{fig:tempo-hier}.  Again, we can access the coverage of the distribution of $\bar{r}_{ij}$; 97.96\%, 100\% and 99.64\% of the 95\% HPD intervals for $\bar{r}_{ij}$ for the tempo, dynamics and timbre metrics, respectively, contain 0.

\subsection{Potential for Variation by Piece}

The $\tau_p$ parameters were introduced into the HMDS model to account for heterogeneous variation in the potential for across-orchestra differences by piece, where this potential for variation is determined by musical characteristics of the specific piece.  The posterior distributions for each piece for $\tau_p$ are given in \autoref{fig:tempo-tau} for the tempo metric, and in  Section 2 of the Supplementary Material \citep{HMDS:supp} for dynamics and timbre.  Across all three metrics, the $\tau_p$ parameters were able to recover the different potentials for across-orchestra variation by piece, and the results correspond to musical expectation based on the score of each piece. 

For example, for the tempo metric, the posterior mean of $\tau_p$ for Symphony No. 6, Movement 1 (No6-01) was among the lowest for all pieces, while Symphony No. 9, Movement 2 (No9-02) was the highest posterior mean (\autoref{fig:tempo-tau}).  This suggests that there is more potential for across-orchestra variation in tempo for No9-02 than for No6-01.  Indeed, this corresponds to features of the musical score for these two pieces \citep{IMSLP}: piece No9-02 has approximately 24 marked tempo changes, while piece No6-01 has no marked tempo changes and only a fermata on the last note of the piece. (We include fermatas and grand pauses in our approximate count of marked tempo changes.  A fermata indicates that a note should be held for a length of time determined by the conductor, while a grand pause indicates a break of complete silence in the piece for an amount of time again determined by the conductor).  Each tempo change denoted in the score is relative, thus allowing for a high level of variation in interpretation between different orchestras on pieces with many marked tempo changes. While tempo changes that are not written in the score can and do occur, the score is the ``ground truth'' and a proxy for the expected potential of variation.  The posterior distributions of the $\tau_p$ parameters for the dynamics and timbre metrics are summarized in Section 2 of the Supplementary Material \citep{HMDS:supp} and also show a correspondence to musical markings in the score of each piece.

\begin{figure}[htbp]
\begin{center}
\includegraphics[width = 0.95\textwidth]{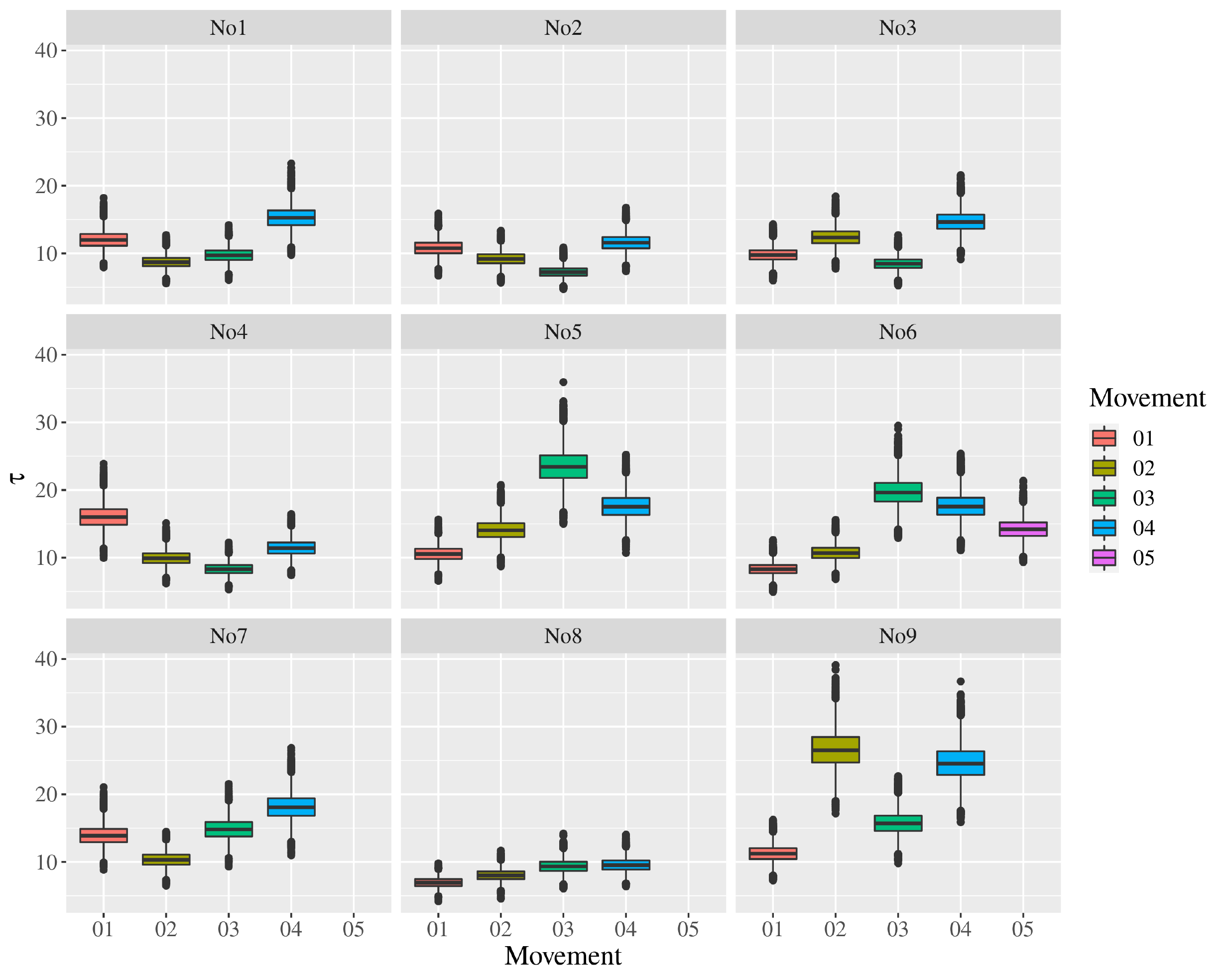}
\caption{Posterior distributions for the $\tau_p$ parameters for the tempo metric, by Beethoven symphony.  The $\tau_p$ parameters were able to recover the variation in the potential for across-orchestra differences by piece, for example, No9-02 as compared to No6-01.}
\label{fig:tempo-tau}
\end{center}
\end{figure}


\subsection{Systematic Differences Between Orchestras}

The primary motivation for this work was to explore systematic variation between orchestras across pieces for various musical metrics, with the secondary goal of relating systematic differences to known characteristics of the orchestra and recording, such as the year of the recording. The latent distance parameters $\{\delta_{ij}: j > i\}$ in the HMDS model capture this systematic variation and are  the main parameters of interest for our application.  From a musical perspective, we do expect systematic differences between orchestras.  For example, the Academy of Ancient Music performs on period instruments, which sound different (in terms of dynamics and timbre) from the modern instruments used by the other orchestras considered here.  Additionally, the musical metrics that we consider can be strongly influenced by the conductor, so we expect the two orchestras under Sir Simon Rattle (Berlin and Vienna) to be similar.  As we will show, the HMDS model is able to recover these expected musical results and to additionally suggest some orchestral similarities that are more surprising.

For each metric, we present a summary of the $\delta_{ij}$'s as a heatmap and a dendrogram formed via hierarchical agglomerative clustering.  For each heatmap, darker colors correspond to smaller values of $\delta_{ij}$, which indicate that orchestra $i$ and orchestra $j$ are more similar to each other.  For each dendrogram, orchestras that are more similar to each other are joined together at a lower height on the dendrogram.    We focus on the posterior mean of each $\delta_{ij}$, as the posterior distribution of each $\delta_{ij}$ is not skewed and fairly symmetric.  Additionally, at each iteration of the Markov chain, the $\delta_{ij}$ values are very close to the values $||X_i - X_j||_2$ and thus satisfy the triangle inequality. 

\subsubsection{Tempo}
\begin{figure}[!htbp] 
\centering
\begin{subfigure}{.5\textwidth}
  \centering
  \includegraphics[width=\linewidth]{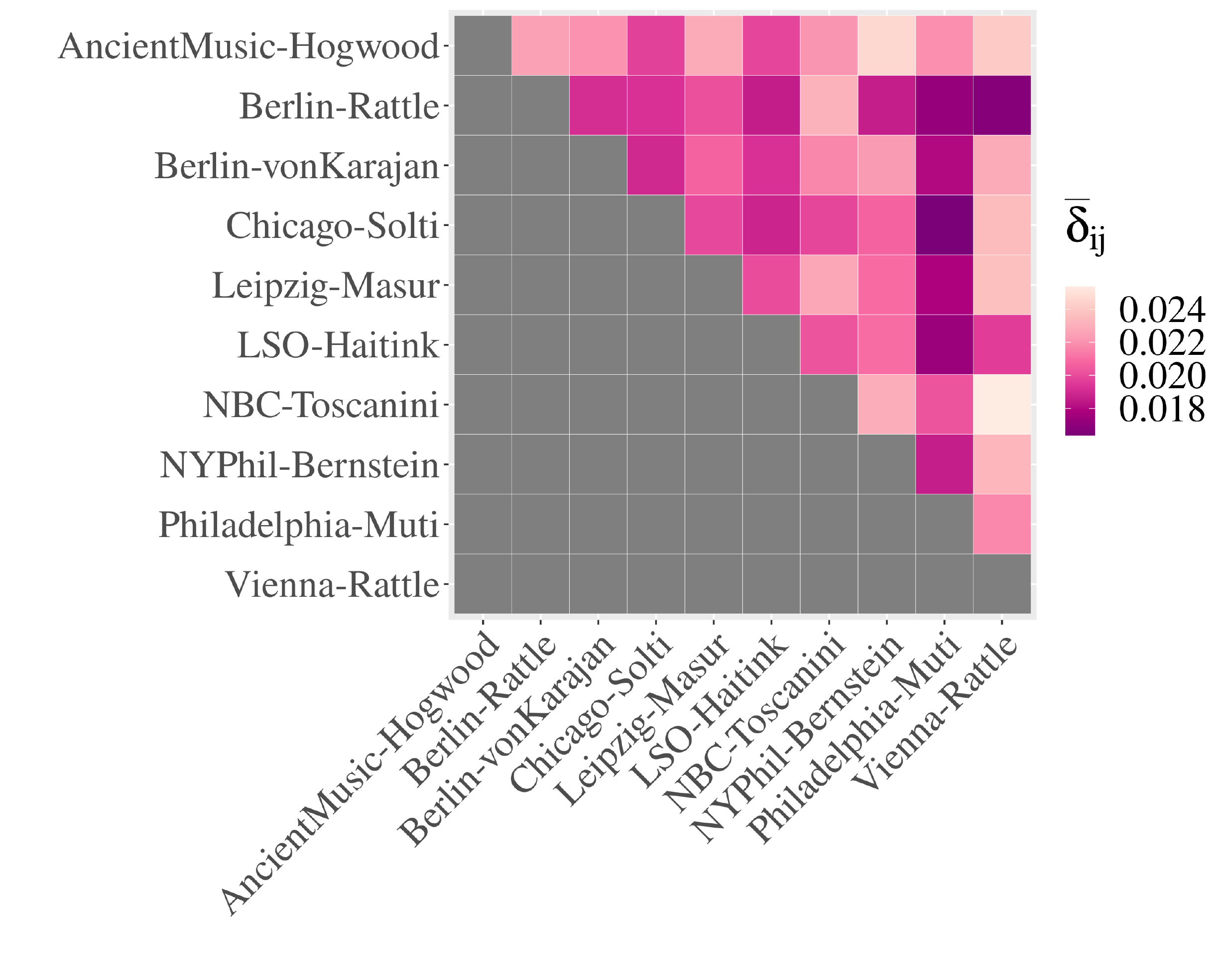}
  \caption{Heatmap}
  \label{fig:tempo-heat}
\end{subfigure}%
\begin{subfigure}{.5\textwidth}
  \centering
  \includegraphics[width=\linewidth]{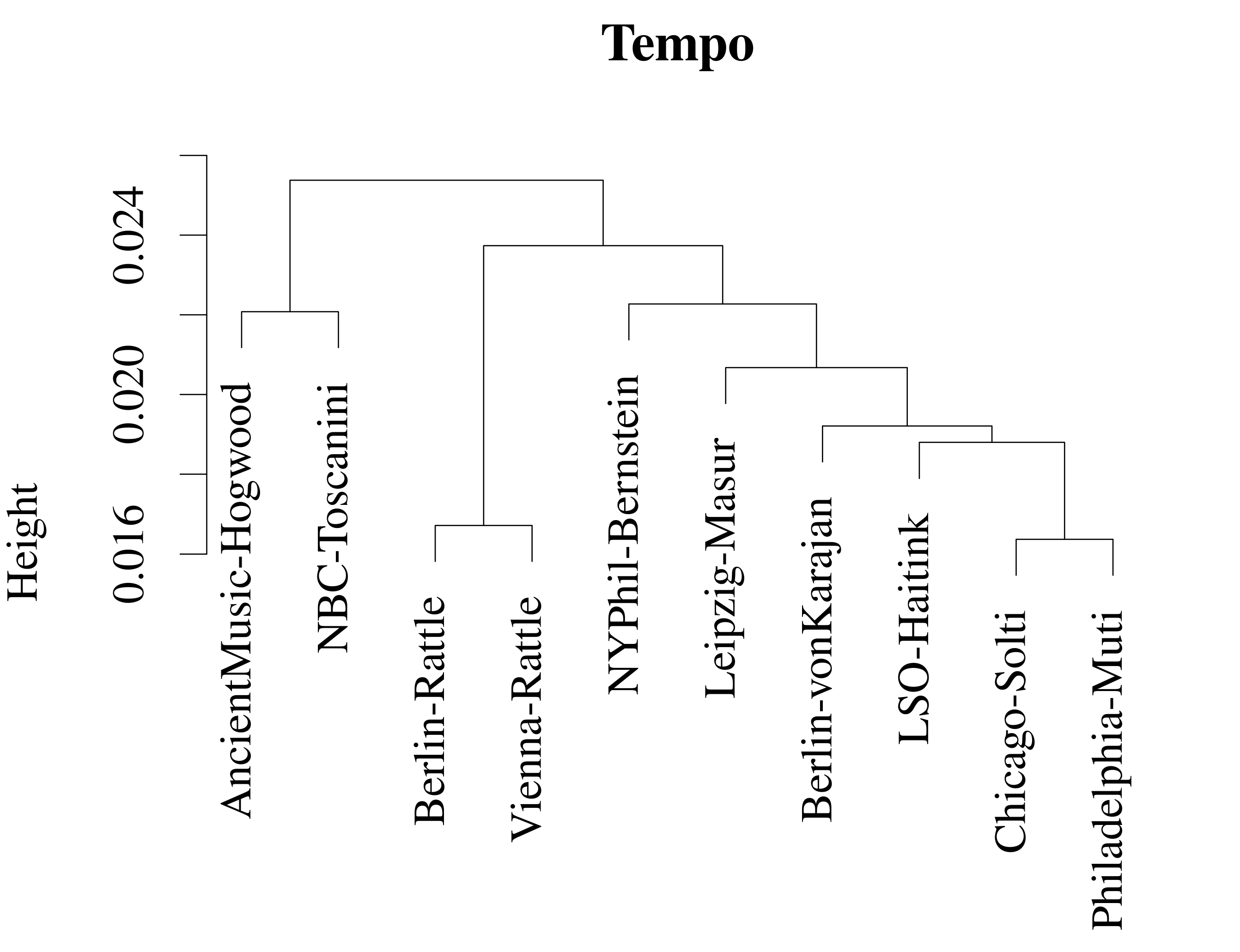}
  \caption{Dendrogram}
  \label{fig:tempo-clust}
\end{subfigure}
\caption{(a) Heatmap of the posterior means of $\delta_{ij}$ and (b) dendrogram found via hierarchical agglomerative clustering on the posterior mean of $\delta_{ij}$ for the tempo metric. The two recordings by Rattle with Vienna and Berlin are very similar. }
\label{fig:tempo}
\end{figure}

A heatmap and dendrogram representing the estimated $\delta_{ij}$'s for the tempo metric are provided in \autoref{fig:tempo-heat} and \autoref{fig:tempo-clust}, respectively.   As expected, the recordings by Sir Simon Rattle with the Vienna Philharmonic and the Berlin Philharmonic are quite similar across pieces.  These recordings were made within 5 years of each other under the same conductor, and the conductor has a large degree of control over the tempo of each piece.  Additionally, the two Rattle recordings are also very similar to the LSO-Haitink recordings in terms of tempo.  The data from these three orchestras are the most recent, in terms of the year in which the recordings were made. However, there does not appear to be any clear evidence of consistent similarity between orchestras by continent.  Somewhat surprisingly, the Academy of Ancient Music and the NBC Symphony Orchestra are  similar to each other, and quite different from the other orchestras.  This result can be interpreted in terms of the style of both orchestras and the scholarship of their conductors.  The Academy of Ancient Music uses scholarship to imitate the way that these pieces would have been performed in Beethoven's time.  The NBC Symphony Orchestra recordings, on the other hand, were conducted by Arturo Toscanini (born in 1867), who was a contemporary of musicians that were alive in Beethoven's time (Beethoven died in 1827) and who studied a style of conducting similar to that of the era when these pieces were composed.  

\subsubsection{Dynamics}
\begin{figure}[!htbp] 
\centering
\begin{subfigure}{.5\textwidth}
  \centering
  \includegraphics[width=\linewidth]{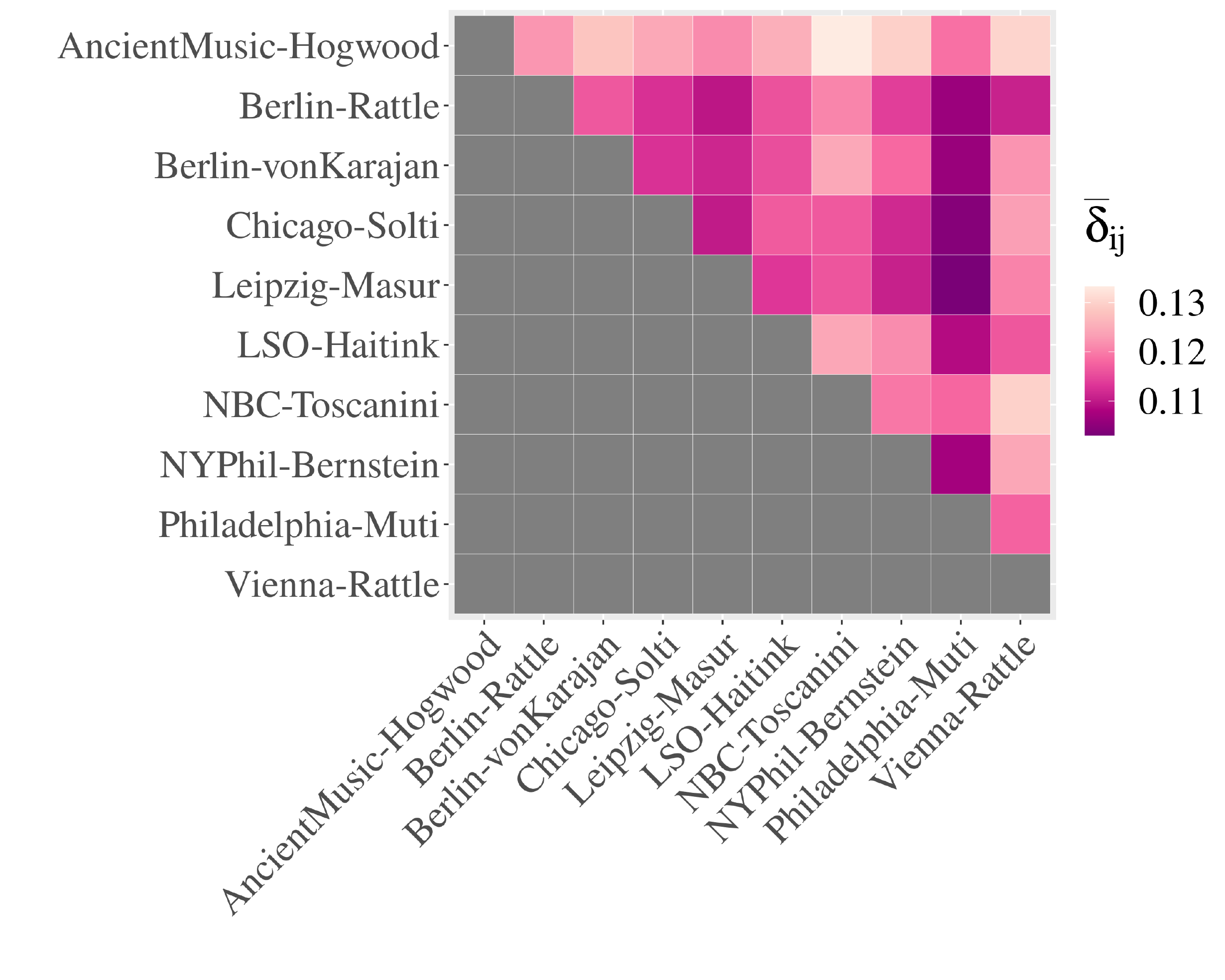}
  \caption{Heatmap}
  \label{fig:volume-heat}
\end{subfigure}%
\begin{subfigure}{.5\textwidth}
  \centering
  \includegraphics[width=\linewidth]{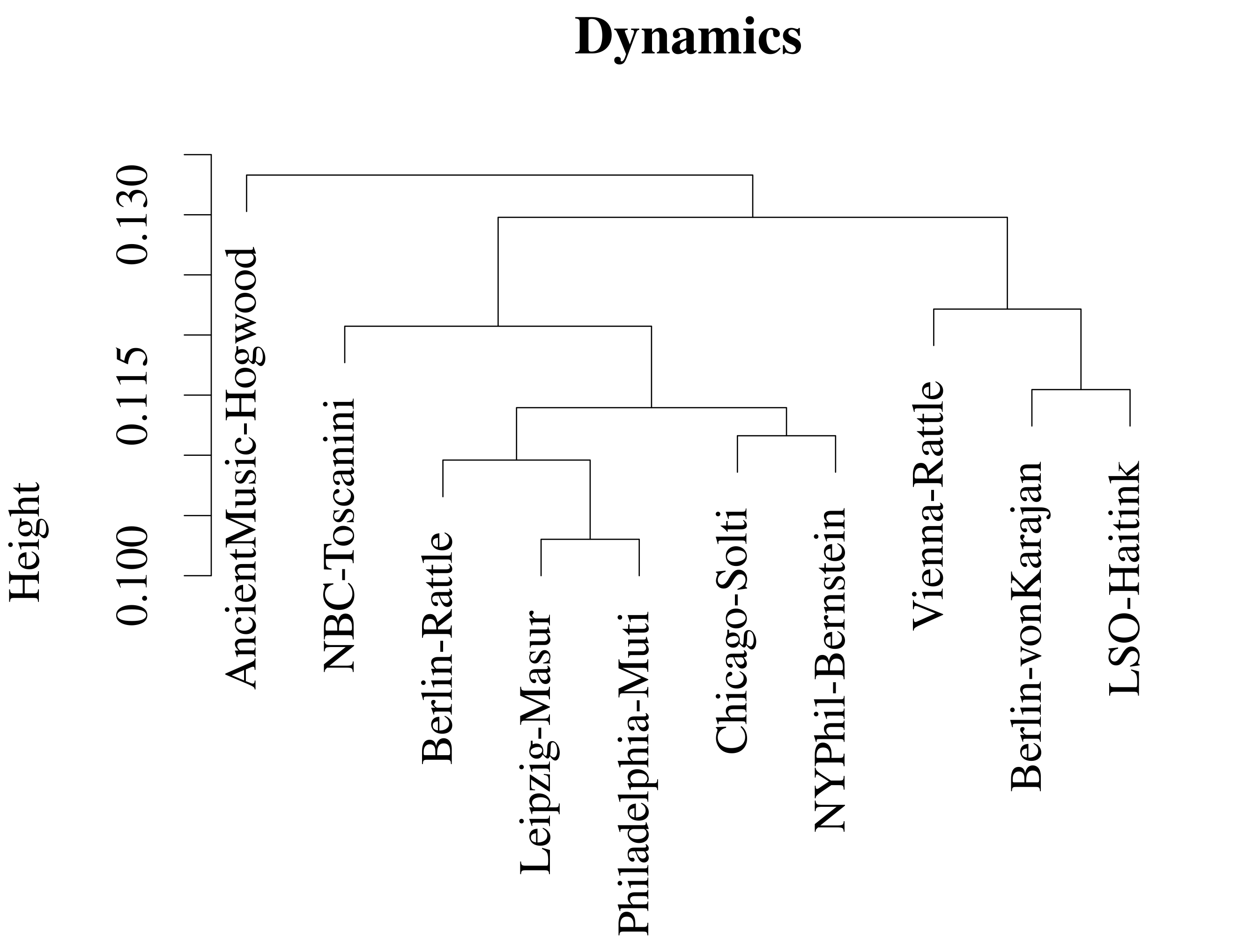}
  \caption{Dendrogram}
  \label{fig:volume-clust}
\end{subfigure}
\caption{(a) Heatmap of the posterior means of $\delta_{ij}$ and (b) dendrogram found via hierarchical agglomerative clustering on the posterior mean of $\delta_{ij}$ for the dynamics metric. The Academy of Ancient Music appears to be an outlier. }
\label{fig:vol}
\end{figure}

The estimates of the $\delta_{ij}$'s for the dynamics metric also confirm prior musical expectations.  In both the heatmap (\autoref{fig:volume-heat}) and the dendrogram (\autoref{fig:volume-clust}) for dynamics, the Academy of Ancient Music appears to be an outlier.  This is expected, as Ancient Music is the only orchestra to record on period instruments, which cannot play as loudly as modern instruments.  Thus, the contrast in the range of dynamics (from loudest to softest volume) is smaller on period instruments than modern orchestral instruments.  NBC also performed on older instruments (from the 1950s), which were not as loud as the modern orchestral instruments used by the remaining orchestras.  Unlike the tempo metric, the dynamics metric shows evidence for some clustering by continent (\autoref{fig:volume-clust}).  Vienna-Rattle, Berlin-von Karajan and LSO-Haitink are very similar, while all of the American orchestras are on the same main branch of the dendrogram. Finally, as shown in \autoref{fig:volume-heat}, there is again evidence for higher similarity between more modern recordings, as Vienna-Rattle, Berlin-Rattle, LSO-Haitink and Philadelphia-Muti are very similar.

\subsubsection{Timbre/Spectral Flatness}
\begin{figure}[!htbp] 
\centering
\begin{subfigure}{.5\textwidth}
  \centering
  \includegraphics[width=\linewidth]{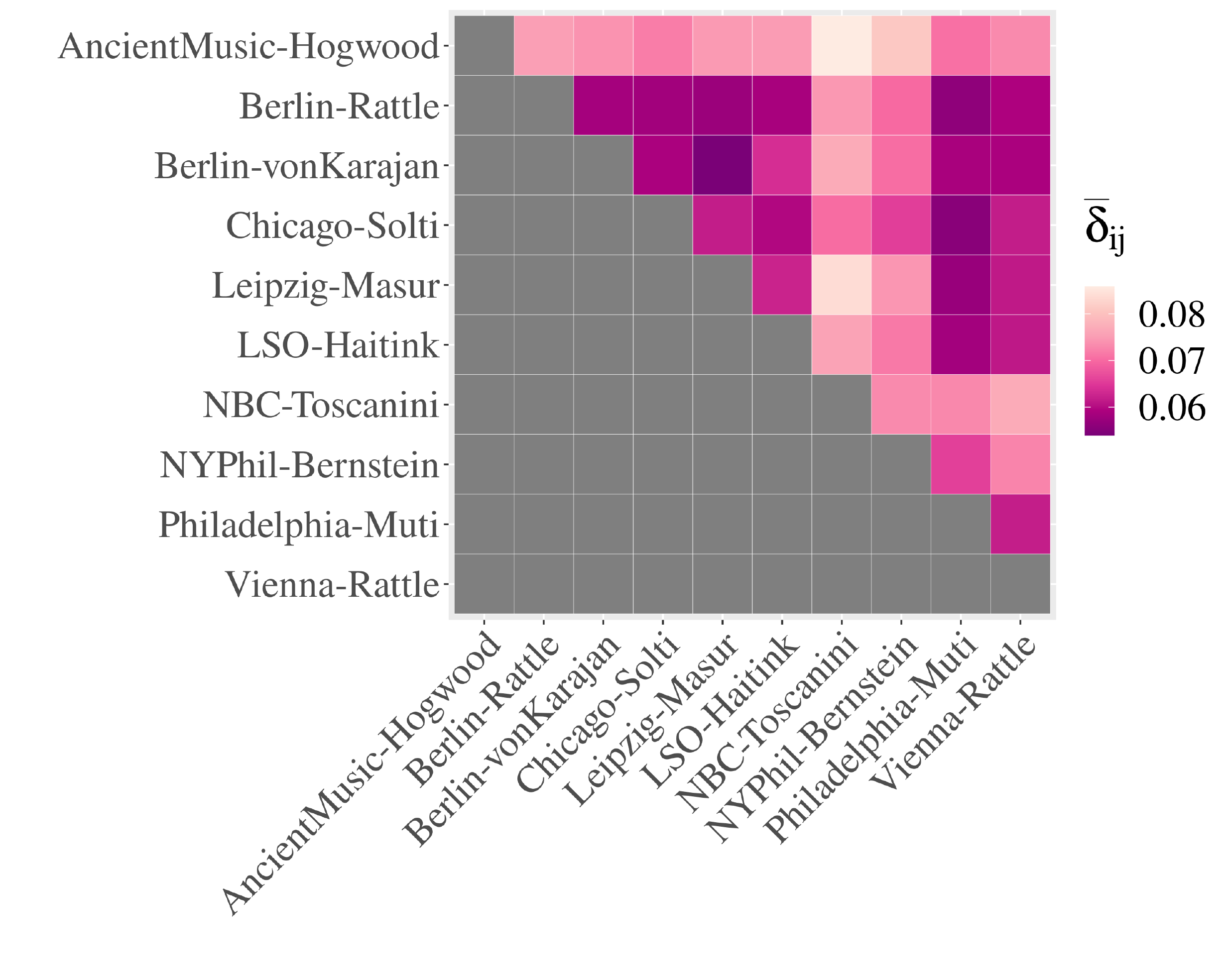}
  \caption{Heatmap}
  \label{fig:SF-heat}
\end{subfigure}%
\begin{subfigure}{.5\textwidth}
  \centering
  \includegraphics[width=\linewidth]{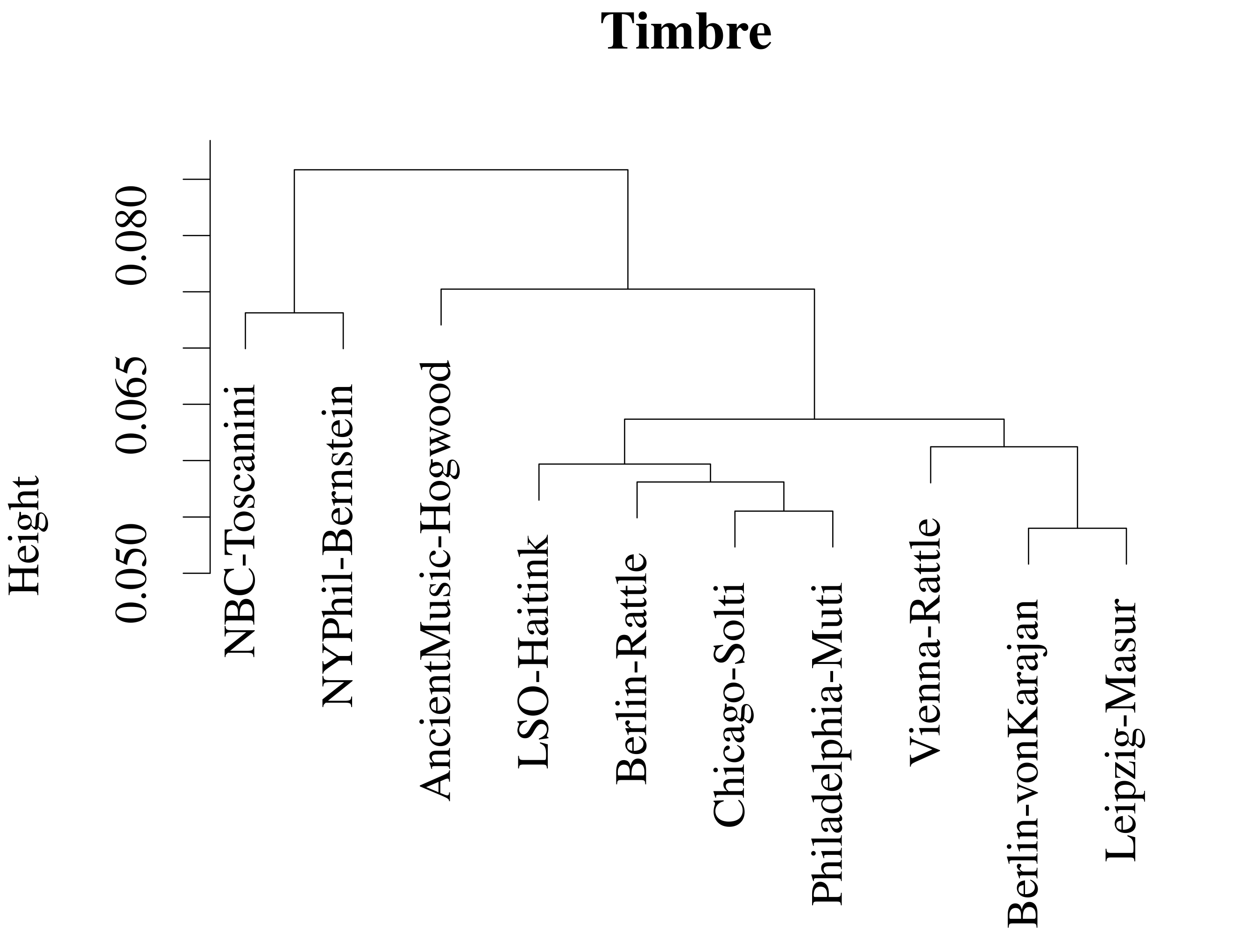}
  \caption{Dendrogram}
  \label{fig:SF-clust}
\end{subfigure}
\caption{(a) Heatmap of the posterior means of $\delta_{ij}$ and (b) dendrogram found via hierarchical agglomerative clustering on the posterior mean of $\delta_{ij}$ for the timbre metric. Newer recordings are more similar to each other than older recordings. }
\label{fig:spec}
\end{figure}

Finally, the analysis of the posterior means of $\delta_{ij}$ for timbre again align with our musical expectation.  Ancient Music is again an outlier for the timbre metric, due to the use of period instruments which have a fundamentally different timbre than modern orchestral instruments (\autoref{fig:SF-heat}).  Additionally, the NBC-Toscanini recordings and NY-Bernstein recordings are similar to each other and different from most of the other orchestras (\autoref{fig:SF-clust}), likely due to specifics of the recording technology at the time these recordings were made.  Again, there is evidence for newer recordings being more similar to each other, as the LSO-Haitink, Berlin-Rattle, Chicago-Solti and Philadelphia-Muti are quite similar to each other in \autoref{fig:SF-clust}.  Finally, there is no consistent evidence for similarity by continent.  While the NY-Bernstein recordings are similar to Chicago and Philadelphia in \autoref{fig:SF-heat},  Philadelphia is also very similar to several European orchestras.

\subsection{Low-Dimensional Embedding}

While the focus of many classical MDS methods is often on dimensionality reduction of the learned embeddings, the primary purpose of HMDS is to share information and identify differences (via the $\delta_{ij}$ parameters) that are consistent across multiple distance matrices, rather than to provide a low-dimensional visualization of the distance data.   For our application, these estimated $\delta_{ij}$'s are not well-represented by points in a Euclidean space with only two dimensions.  For example, a principal components analysis on the $X_i$ vectors indicates that all of the principal components are roughly equal in importance (Section 2 of the Supplementary Material \citep{HMDS:supp}).

\begin{figure}
\begin{center}
\includegraphics[width = 0.7\textwidth]{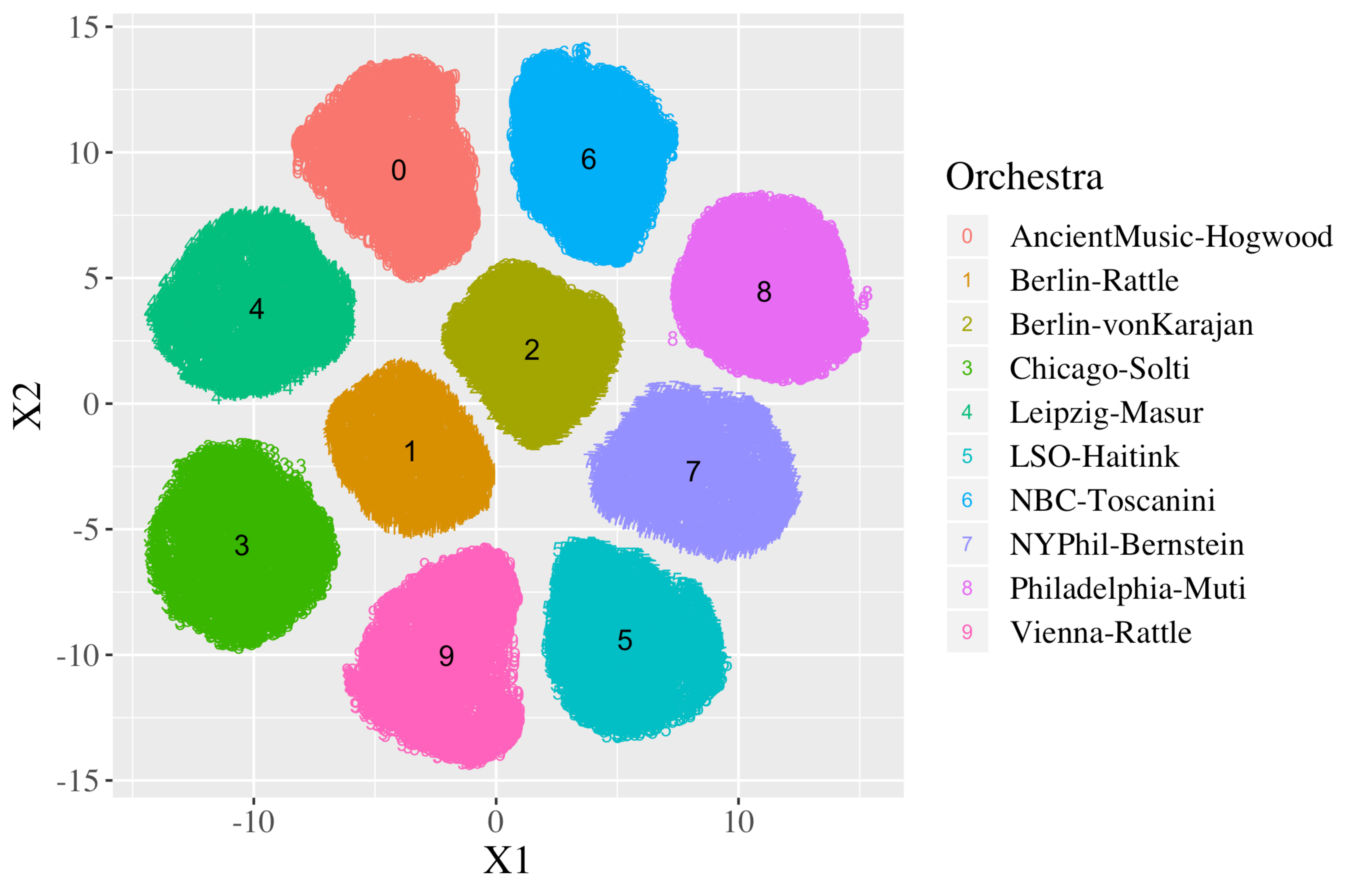}
\caption{A t-SNE embedding for the posterior distribution of the $X_i$'s for the tempo metric. Each point represents an iteration of the Markov chain for $X_i$ embedded in a two-dimensional space.  The similarity between orchestras corresponds to the results for the hierarchical clustering of the posterior means for $\delta_{ij}$.  Additionally, the $X_i$ vectors separate well by orchestra.}
\label{fig:tsne}
\end{center}
\end{figure}

We instead use a non-linear dimensionality reduction approach to visualize the $X_i$ vectors.  The Procrustes transformation is first used to align each sampled $X_i$ to the posterior mean, for each metric individually.  The aligned $X_i$ vectors lie in 9-dimensional space and a t-SNE \citep{tSNE:2008} embedding is then used to visualize the $X_i$ vectors in 2-dimensional space.  The t-SNE embedding is able to retain local and global structure in the data through the dimensionality reduction by keeping points that are close in the higher dimensional space close in the lower dimensional space \citep{tSNE:2008}.  The posterior distributions for the embedded $X_i$ vectors separate well by orchestra and are arranged similarly to the findings with the hierarchical clustering for $\delta_{ij}$ above (\autoref{fig:tsne}).  For example, for the tempo metric, the $X_i$ vectors for Berlin-Rattle and Vienna-Rattle are quite close to each other, indicating that the recordings by these two orchestras are similar in terms of tempo.  The $X_i$ vectors for NBC and the Academy of Ancient Music are also close to each other, again indicating similarity in terms of tempo for these two orchestras.

\section{Conclusions and Future Work}\label{sec:conc}

In order to quantify systematic differences between orchestras, we developed a hierarchical, model-based approach to multidimensional scaling.  This method generalized the BMDS model of \citet{Oh-Raftery:2001} in several ways, including the extension to modeling heterogeneous replicate distance matrices. We applied HMDS to the comparison of different orchestras across the Beethoven symphonies. The proposed HMDS model was successful in uncovering systematic differences between orchestras across pieces and the $\tau_p$ parameters were able to capture variation in the potential for across-orchestra differences. The overall analysis of the posterior means for $\delta_{ij}$, the systematic dissimilarity between orchestras $i$ and $j$, for the three musical metrics yielded some expected and surprising results.  As expected, the recordings by Vienna-Rattle and Berlin-Rattle were found to be very similar across all three musical metrics, but especially tempo, which the conductor has a large influence on.  Additionally, the Academy of Ancient Music was confirmed as an outlier in terms of dynamics and timbre, due to their use of period instruments.  Surprisingly, we found that NBC and the Academy of Ancient Music were most similar to each other in terms of tempo, and this might be attributed to their adherence to artistic styles from Beethoven's time. 

However, a few other interesting results also emerged.  Across all three of the musical metrics,  Philadelphia-Muti and Leipzig-Masur were quite similar.  This was not an obvious result, as Riccardo Muti and Kurt Masur had different conducting backgrounds and experiences.  Additionally, the Philadelphia Orchestra and the Leipzig Gewandhaus Orchestra are both among the most well known American and European orchestras, respectively, with very different histories.  Philadelphia is known for the ``Philadelphia Sound'', developed under Leopold Stokowski in the early 1900s, while Leipzig has a long and storied history of conductors, including Felix Mendelssohn.

Finally, the evidence for more similarity across tempo, dynamics and timbre in the more recent recordings was not necessarily expected a priori \citep{Liebman:2012}.  While some of this is due to changes in recording technology, and the fact that Rattle conducted both the Vienna and Berlin Philharmonics here, this result suggests that there may be less variation among newer recordings by different orchestras, as compared to older recordings. 

It is important to note that in terms of the musical application of interest, tempo was the only metric considered that was independent of the recording technology used for each orchestra's recordings, and from that perspective, was the most indicative of artistic differences between orchestras.  Dynamics and timbre both depend on the specifics of recording technology, and are also related to each other, as louder instruments will have different timbres than softer overall instruments. Additionally, it should be noted that the different orchestras considered performed on different brands and makes of instruments, which also contributed to the differences in dynamics and timbre.  Thus, future work could include additional, and more nuanced, musical metrics to further support our findings here. Alternative musical metrics would also be important to explore in the case of other composers, particularly composers with less potential for variation along the musical metrics considered here.

Future extensions to the proposed HMDS model could consider a hierarchical extension to share information across metrics and learn an overall distance for each orchestra pair.  Other extensions could include exploring constraints on the parameter space for the $X_i$ vectors to allow for identifiability of these parameters, such as in \citet{Bakker:2013}.  Additionally, while HMDS is designed for the quantitative comparison of orchestral audio recordings, the model is flexible enough to be of use in several different application domains with similar data structures.  Of critical importance here, is that comparing entity $i$ on replication $p$ to entity $j$ on replication $p'$ does not make sense, meaning that approaches similar to distance-based ANOVA \citep{Anderson:2001, McArdle-Anderson:2001} cannot be used.  Thus, HMDS is especially applicable for medical settings, for example, where there are many different types of measurements made (replications $p$), but where each type of measurement cannot be compared directly to another, and instead, the question of interest is the existence of systematic similarities between patients or individuals across measurement types.  Furthermore, the $\tau_p$ parameters allow for different potentials of variation for each measurement type, a fact that may be especially relevant in many medical applications, as well.


\section*{Acknowledgements}
The authors would like to thank Mike Kris for helpful discussions about the interpretation of results in a musical context.  The authors would also like to thank the four anonymous referees and the Editor for their constructive and helpful comments that improved the quality of this paper.

\begin{supplement} 
\stitle{Supplement to ``Hierarchical multidimensional scaling for the comparison of musical performance styles''}
\slink[doi]{10.1214/20-AOAS1391SUPPA}
\sdatatype{.pdf}
\sdescription{We provide details of one possible Markov chain Monte Carlo algorithm for inference in the HMDS model, and additional pre-processing details, model diagnostics and results.}
\end{supplement}
\begin{supplement}
\stitle{Source code and data for ``Hierarchical multidimensional scaling for the comparison of musical performance styles''}
\slink[doi]{10.1214/20-AOAS1391SUPPB}
\sdatatype{.zip}
\sdescription{R code for the proposed HMDS model and data for the comparison of Beethoven symphonies across 10 orchestras and multiple musical metrics.}
\end{supplement}



\bibliographystyle{imsart-nameyear} 
\bibliography{refs}
\break\newpage

\end{document}